\documentclass[useAMS,usenatbib]{mn2e}

\usepackage{enumerate}
\usepackage{graphicx}
\usepackage[usenames]{color}

\newcommand{\simgt}{\lower.5ex\hbox{$\;\buildrel>\over\sim\;$}}
\newcommand{\simlt}{\lower.5ex\hbox{$\;\buildrel<\over\sim\;$}}

\newcommand{\msun}{\ensuremath{M_\odot}}
\newcommand{\msunyr}{$M_\odot$\,yr$^{-1}$}

\newcommand{\mstar}{\ensuremath{M_{\rm star}}}
\newcommand{\psf}{$P_\mathit{sf}$}

\newcommand{\hii}{\rm H{\sc ii}}

\newcommand{\ha}{H$\alpha$}
\newcommand{\hb}{H$\beta$}
\newcommand{\bra}{Br$\alpha$}

\newcommand{\htwo}{H$_2$}
\newcommand{\xco}{$X_{\rm CO}$}

\newcommand{\logoh}{12$+\log$(O/H)}

\newcommand{\sbs}{SBS\,0335$-$052}
\newcommand{\izw}{I\,Zw\,18}
\newcommand{\spit}{{\it Spitzer}}
\newcommand{\oiii}{\rm O{\sc iii}}

\newcommand{\aap}{A\&A}
\newcommand{\apj} {ApJ}
\newcommand{\apjl} {ApJL}
\newcommand{\apjs} {ApJS}
\newcommand{\aj} {AJ}
\newcommand{\araa} {ARA\&A}
\newcommand{\mnras}{MNRAS}
\newcommand{\nat}{Nature}

\newcommand{\pasp}{PASP}

\title[Scaling relations of metal-poor starbursts. II]{Scaling relations of
metallicity, stellar mass, and star formation rate in metal-poor starbursts: II. Theoretical models}

\author[L.~Magrini et al.]{Laura Magrini$^{1}$\thanks{E-mail: laura@arcetri.astro.it},
Leslie Hunt$^{1}$, Daniele Galli$^{1}$, Raffaella Schneider$^{2}$, Simone Bianchi$^{1}$,
\newauthor 
Roberto Maiolino$^{3}$,
Donatella Romano$^{4}$,
Monica Tosi$^{4}$, 
Rosa Valiante$^{2}$\\ 
$^{1}$INAF/Osservatorio Astrofisico di Arcetri, Largo Enrico Fermi 5, 50125 Firenze, Italy\\
$^{2}$INAF/Osservatorio Astronomico di Roma, Via di Frascati 33, 00040 Monteporzio, Italy\\
$^{3}$Cavendish Laboratory, University of Cambridge, 19 JJ Thomson Avenue, Cambridge CB3 0HE, UK\\
$^{4}$INAF/Osservatorio Astronomico di Bologna, Via Ranzani 1, 40127 Bologna, Italy}
\begin{document}

\date{}

\pagerange{\pageref{firstpage}--\pageref{lastpage}} \pubyear{2012}

\maketitle

\label{firstpage}

\begin{abstract} 
{Scaling relations of metallicity (O/H), star formation
rate (SFR), and stellar mass (\mstar) give important insight on galaxy
evolution.  They are obeyed by most galaxies in the Local Universe and
also at high redshift.  In a companion paper, we compiled a sample
of $\sim$1100 galaxies from redshift 0 to $\ga3$, spanning almost two
orders of magnitude in metal abundance, a factor of $\sim10^6$ in SFR,
and of $\sim10^5$ in stellar mass.  We have 
characterized empirically the star-formation ``main sequence'' (SFMS)
and the mass-metallicity relation (MZR) for this sample, and also
identified a class of low-metallicity starbursts, rare locally but more common in the
distant universe. These galaxies deviate significantly from the main scaling
relations, with  high SFR and low metal content for a given \mstar.
In this paper, we model the scaling relations and explain these
deviations from them with a set of multi-phase chemical evolution
models based on the idea that, independently of redshift,
initial physical conditions in a galaxy's evolutionary history can
dictate its location in the scaling relations.  Our models are able to
successfully reproduce the O/H, \mstar, and SFR scaling relations up to
$z\simgt3$, and
also successfully predict the molecular cloud fraction as a function of
stellar mass.  These results suggest that the scaling relations are defined
by different modes of star formation: an ``active'' starburst mode, more
common at high redshift, and a quiescent ``passive'' mode that is predominant
locally and governs the main trends.
} 
\end{abstract}


\section{Introduction} \label{sec:intro}

Scaling relations of metallicity O/H, stellar mass \mstar, and star-formation rate (SFR)
are powerful tools for probing galaxy evolution across cosmic time.
The star-formation ``main sequence'' (SFMS) connects current star formation
with a galaxy's stellar mass \citep{noeske07,salim07,schiminovich07,bauer11},
and the mass-metallicity relation (MZR) links 
stellar mass with metal content \citep{tremonti04}.
Recently, \citet{mannucci10} found that SFR influences a galaxy's position
in the MZR, and introduced the ``Fundamental Metallicity Relation'' (FMR), which
reduces the scatter in their local sample of galaxies to $\sim$0.06\,dex 
\citep[see also][]{lara10}. 

There is, however, evidence that galaxies outside the Local Universe ($z\ga1$) 
do not follow the same SFMS and MZR.
The FMR successfully fits Lyman Break Galaxies (LBGs) up to $z\la2$, but fails
for higher redshifts \citep{mannucci10}.
The SFMS has similar slopes with increasing redshift but larger offsets \citep{bauer11};
for a given stellar mass, a more distant galaxy has a higher SFR.
Starbursts at all redshifts tend to lie above the SFMS, with high
specific SFRs (sSFRs), and too much
luminosity (or stellar mass) for their metallicity
\citep[e.g.,][]{hoyos05,rosario08,salzer09,peeples09}. 
These high-SFR outliers have been attributed to merger-induced episodes of
star formation that tend to be short-lived and intense \citep[e.g.,][]{rodighiero11}, 
while the SFMS is followed by more passively evolving galaxies that form stars 
over longer times.

Here we propose that mergers may not be directly responsible for producing
the starbursts that depart from the general SFR-mass-metallicity relations.
Instead, independently of their cause,
different modes of star formation arising from different initial sizes and densities, 
may be driving the deviations.
{\em Active}
star formation that takes place in compact regions with dense gas has
distinct properties from a more {\em passive} mode in larger, more tenuous regions;
in the former, dynamical times are shorter, infrared luminosities and SFRs are higher, 
and molecular fractions are more elevated \citep{hirashita04}.
In the Local Universe, 
compact sizes and high-density gas characterize metal-poor starbursts
which occur in dwarf galaxies with high SFRs \citep{hunt09}.
At high redshift, starbursts also tend to be compact in size
\citep[$\la 4$\,kpc in diameter,][]{tacconi06,tacconi08,toft09}\footnote{However,
quiescent galaxies at $z\sim2$ tend to be more compact than starbursts as the
same redshift, implying that those galaxies were formed by compact nuclear starbursts
at even higher redshifts, $z\sim3-4$.}.
Compactness also shapes spectral energy distributions causing them to peak
at shorter wavelengths, because the dust
heated by a compact starburst tends to be warmer
\citep{chanial07,groves08,melbourne09}.

\begin{figure}
\hbox{
\includegraphics[width=\linewidth,bb=18 144 592 650]{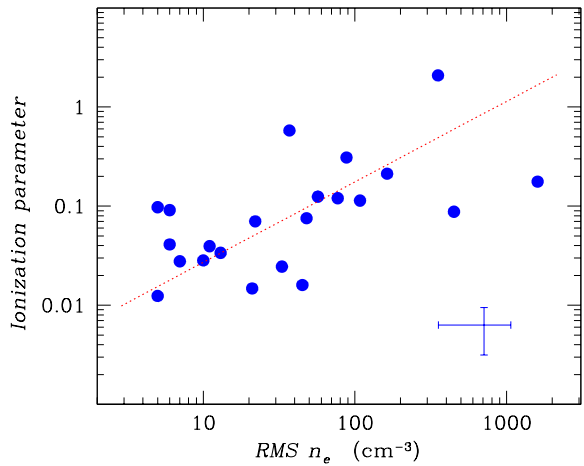} 
}
\caption{Ionization parameter vs. root-mean-square electron density
of \hii\ regions in a sample of BCDs from \citet{hunt09}.
The typical error bars are shown in the lower right corner.
The dotted line shows the best-fit correlation.
}
\label{fig:ion}
\end{figure}

Active star formation is expected to also be characterized by high ionization
parameters because of the compact and dense nature of the active SF complexes.
Figure~\ref{fig:ion} shows the ionization parameters $U$ of a sample of 
low-metallicity Blue Compact Dwarf galaxies (BCDs) taken from \citet{hunt09}.
The ionization parameter
of the \hii\ regions increases with ionized gas density $n_e$; for this
sample the significance of the correlation is $>$99\%.
Because the size of the \hii\ region is inversely correlated with density 
\citep[e.g.,][]{hunt09}, $U$ depends almost linearly on $n_e$.
Anomalous \hii-region excitation is frequently found
in some high-redshift galaxy populations 
\citep[e.g.,][]{hammer97,liu08,brinchmann08,erb10}.
Excluding a contribution from a weak active galactic nucleus,
it is likely that this excitation results from the
extreme physical conditions of the {\em active} SF mode,
with dense ionized gas and high SFR surface densities \citep{liu08}.
Such properties naturally arise in dense and compact \hii\
regions with particularly hard radiation fields.
Thus, the anomalous excitation and high ionization parameters
observed in high-redshift samples may also be signatures of 
active starbursts which occur in a chemically unenriched interstellar 
medium (ISM). 

Besides compact, dense star-forming regions,
and possibly high-ionization parameters for the nebular gas,
there are further, probably related,
properties shared by some galaxy populations in the nearby universe
and those at high redshift.
As discussed in \citet[][hereafter Paper~I]{hunt12}, some metal-poor BCDs 
deviate significantly from the SFMS and the MZR; they are
characterized
by high SFRs, high sSFRs, and are situated in the SFMS and the MZR in the same
region as the $z=3$ LBGs. 
For a given (sub-solar) metallicity,
these galaxies can have an excess of stellar mass of two orders of magnitude,
and a factor of $\sim10$ excess SFR for a given mass.
Some of the ``Green Peas'' and Luminous Compact Galaxies (LCGs) at $z\sim0.1-0.4$
selected by \citet{cardamone09} and \citet{izotov11} also lie in the same
regions of the SFMS and the MZR and show similar excesses.
Similar physical conditions seem to occur at all redshifts, but
different galaxy populations are observed at
different redshifts because of how they are selected. 
Since extreme starbursts are rare locally, but more common at high redshift, 
it could be that different modes of star formation 
drive the scaling relations of stellar mass, metallicity, and SFR we observe.

In this paper, we present semi-analytical models which 
are aimed at understanding 
how different modes of star formation affect the SFMS and MZR.
In particular,
we focus on metal-poor galaxies with high SFRs, which are 
generally more massive than would be expected given their metal abundance.
We develop three
families of initial conditions for star formation and compare model 
predictions with several samples comprising $\la$ 1100 nearby metal-poor
dwarf galaxies, intermediate redshift and high-redshift galaxy populations 
(see Paper~I).
The initial conditions are designed to embrace the active/passive 
(compact$+$dense/extended$+$diffuse) modes of star formation and thus
distinguish the {\em active} starburst modes from more {\em passive} quiescent ones.
The questions we aim to address are:
\begin{enumerate}[(a)] 
\item Why can both high and low SFRs be observed at the same metallicity?
\item What are the specific conditions which distinguish starbursts 
(active SF mode) from quiescent star formation (passive mode)?
\item Is there a connection between active dwarfs and high-z starbursts
in terms of their scaling relations (SFMS, MZR, FMR)?
\end{enumerate}
The paper is organized as follows:
in Sect.~\ref{sec:sample} we briefly describe the samples
we have compiled and how we calculate the stellar mass.
The sample is more completely described in Paper~I.
In Sect.~\ref{sec:models}, we discuss our semi-analytical models
which consider different sets of initial conditions in order to 
mimic the starburst and quiescent modes of star formation.
Comparisons of model predictions with observations are given
in Sect.~\ref{sec:comparison}, and 
Sect.~\ref{sec:gas} discusses the relation of the gas-to-stellar
mass ratio and metal abundance. 
In Sect.~\ref{sec:discussion} we describe
possible similarities between nearby and distant starbursts,
and how the scaling relations we observe and successfully model
can shed light on how galaxies evolve. 

\section{The sample and the observables} \label{sec:sample}

Our sample was compiled on the basis of three ``pseudo-observational'' 
parameters\footnote{We say ``pseudo-observational" because these variables are {\it derived}
from observations rather than being directly observed.
Nevertheless, for simplicity, we will call O/H, SFR, and \mstar\ ``observables''.}:
SFR, metal abundance [as defined by the nebular oxygen abundance, \logoh], 
and stellar mass, \mstar.
We included only galaxies with these data either already published, or
for which we could derive the quantities ourselves (in particular the stellar masses).
These criteria were met by galaxies from five samples at $z\sim0$:
21 nearby dwarf irregular galaxies (dIrr) published by \citet[][hereafter ``dIrr'']{lee06},
129 of the 11\,Mpc distance-limited sample of nearby galaxies \citep[11HUGS, LVL:][]{kennicutt08},
and 89 BCDs in the samples presented by \citet{engelbracht08}, \citet{fumagalli10}, 
and \citet{hunt10}.
Seven galaxy samples at higher redshifts also have the necessary data:
the ``Green Pea'' compact galaxies identified by the Galaxy Zoo 
team at $z\sim0.1-0.3$ \citep{cardamone09},
the LCG sample at $z\sim0.1-0.6$ selected by \citet{izotov11},
LBGs at $z\sim1$ \citep{shapley05a}, $z\sim2$ \citep{shapley04,erb06}, and
$z\sim3$ \citep{maiolino08,mannucci09}.
When stellar masses were not available in the literature,
we determined stellar masses from IRAC photometry as discussed below, and in detail
in Paper~I.

Altogether, we consider in the analysis 1070 galaxies from $z\sim0$ to $z\sim3$
with the requisite three quantities of SFR, \logoh, and \mstar.
Except for the galaxies with $z\ga1$ \citep{shapley04,shapley05a,erb06,maiolino08,mannucci09}
and a few of the LVL galaxies \citep{marble10}, metallicities for all galaxies
are determined by the ``direct'' (electron temperature) method \citep[e.g.,][]{izotov07}
since the samples are defined by requiring detections of \oiii$\lambda$4363.  
SFRs are derived from \ha\ luminosities corrected for extinction; in a few BCDs,
the IR luminosity gave higher SFRs, so we used the higher value. 
SFRs using the \citet{kennicutt98} conversion are reported to the \citet{chabrier03} Initial
Mass Function (IMF).
Paper I gives a detailed description of the sample.

We calculated the stellar masses for the local samples (dIrr, BCDs, 11HUGS/LVL)
from \spit/IRAC observations at 4.5\,\micron\ (and when available also 3.6\,\micron).
We based our method on that used by \citet{lee06} who analyzed 
sub-solar metallicity models from \citet{bell01} to
derive stellar mass-to-light (M/L) as a linear function of $B-K$.
The advantage of using 4.5\,\micron\ to measure stellar masses
is that it minimizes variations in the M/L ratio \citep[e.g.,][]{jun08}, 
because the ratio depends only weakly on age and metallicity, and dust
extinction is negligible.

However, before calculating the masses, we subtracted the nebular continuum
and recombination line emission from the broadband 4.5\,\micron\ photometry.
Nebular emission in metal-poor starbursts, both in the continuum and lines,
can significantly affect photometry
\citep[e.g.,][]{reines10,atek11}, and
thus have a potentially strong impact on the inferred stellar masses.
This can be especially important in the 4.5\,\micron\ band \citep{smith09}.
Hence, before applying the formalism of \citet{lee06}, from the SFR
we have inferred the strength of \bra\ emission and the nebular continuum in 
the 4.5\,\micron\ IRAC band, and subtracted it from the observed flux.
Then, following \citet{lee06}, 4.5\,\micron\ luminosity was converted to stellar mass with 
a M/L corrected for different ages and metallicities with a color correction $B-$[4.5].
All stellar masses are scaled to the \citet{chabrier03} IMF. 

The final combined sample covers a $\sim10^5$ range in stellar mass,
a factor of $\sim10^6$ in SFR, and 2 orders of magnitude in oxygen abundance.
The metallicities are generally (formally) quite accurate, with uncertainties on the
direct method of $\simlt$0.05\,dex, but the strong-line method and its comparison
with the direct method of determining abundances from electron temperatures
are compromised by differences in calibration that can be quite large
\citep[e.g.,][see also Paper~I]{kewley08}.
The SFRs suffer from the typical uncertainties of H$\alpha$ measurements
and the uncertain extinction corrections, and the uncertainty is expected to
be $\simlt$30\%.
The uncertainties on the stellar masses are typically 0.1\,dex (30\%), but can
be as high as 50-70\% when there are unusual colors.
In any case, the large dynamic range of our combined sample should overwhelm any 
uncertainties on individual galaxy parameters.
More details are given in Paper~I.

\subsection{Metal-poor starbursts}

Since we are interested in deviations from the scaling relations,
particularly those of galaxies with high SFR, relatively high mass, and low metallicity,
in Paper~I we quantified a ``low-metallicity starburst".
Because deviations from the MZR become significant at abundances at or below
\logoh$\sim$8.0, we consider this as a threshold below which a galaxy can be
considered ''low-metallicity''.
Deviations from the SFMS and the MZR become noticeable at SFRs of $\sim0.6$\,\msunyr.
Hence we loosely define low-metallicity starburst (LMS)
as a galaxy with a SFR $\geq$0.6\,\msunyr and with \logoh$\leq$8.0.
This is a rather arbitrary label, but one that enables us to identify the galaxies
most likely to deviate from the scaling relations (see Paper~I).
Among the various samples, the LMSs are relatively rare: 2\% of the LVL fall into
this category; $\sim$11\% of the BCDs; 22\% of the LCGs; and $\sim$27\% of
the high-$z$ LBGs. 
Clearly selection effects (e.g., emission line surveys for the BCDs, emission-line
flux for the LCGs) make it more
likely for a galaxy to be a LMS, but there is some indication that LMSs are more
common at high redshift.

\bigskip

\section{Theoretical Models } \label{sec:models}

Here we develop the models with which we explore the idea that
particular initial physical conditions in a galaxy's evolutionary
history can change its location in the scaling relations independently
of redshift.  In particular, we have devised three sets of starting
parameters which dictate the subsequent evolution of the galaxy, and
which encompass the observed distinctions between ``active'' and
``passive'' star formation (see Sect. \ref{sec:intro}).  The models and
the different initial conditions are described below.

Although our models have been shown to realistically approximate the chemical
evolution in nearby galaxies \citep[e.g.,][]{molla96,molla97,molla05,magrini07},
in those models gas accretion has been considered.
Here, as a first approximation, we use a closed box in which
neither infall of pristine gas nor outflow of metal-enriched material are contemplated.
Despite these significant limitations, which will be addressed in future work,
we will show below that our models are remarkably successful in predicting
the details of the SFMS and MZR scaling relations.
They also are consistent with the observed gas fractions both locally and at
high redshift. 

\subsection{The multiphase chemical evolution model}\label{mice}
 
The models developed for the present work have been renamed as MICE
(\underline{M}ultiphase \underline{I}nterstellar medium
\underline{C}hemical \underline{E}volution) models. MICE models adopted
here are a generalization of the multi-phase model by
\citet{ferrini92}, originally built for the solar neighborhood, and
subsequently extended to the entire Galaxy \citep{ferrini94}, and to
other disk galaxies 
\citep[e.g.,][]{molla96,molla97,molla05,magrini07,magrini09}.
We refer to
those papers for a detailed description of the general formalism. 
The main differences of the models presented here relative to those used to 
describe the chemical evolution of disk galaxies are:  
{\em i}) MICE models are one-zone models, with a
single galactic component and without infall or outflow processes; {\em
ii}) no radial variations are considered, so that a uniform evolution is
assumed for the whole galaxy.

Each galaxy is composed of four components: diffuse gas ($g$\/), clouds ($c$\/), stars
($s$\/) and stellar remnants ($r$\/).  At $t=0$ all the mass of the
galaxy is in the form of diffuse gas; subsequently, the mass fractions are
modified by several conversion processes.  The main processes
considered in the model are:  ({\em i}\/) conversion of diffuse gas
into clouds; ({\em ii}\/) formation of stars from clouds and ({\em
iii}\/) disruption of clouds by previous generations of massive stars;
({\em iv}\/) evolution of stars into remnants and return of a fraction
of their mass to the diffuse gas.

We describe in the following the main equations of the model. 
Stars form  by cloud-cloud collisions with a rate \psf\
and by the interactions of massive stars with clouds with rate $P_i$
(induced star formation),
\begin{equation}
\frac{{\rm d}f_s}{{\rm d}t}=P_\mathit{sf} f_c^2+P_{i}f_sf_c-Df_s,
\label{eqn:eqstars}
\end{equation}
where $f_s(t)$ and $f_c(t)$ are the mass fraction in stars and clouds, 
respectively, and
$D$ is the stellar death rate. The quadratic dependence of the SFR
on the mass of clouds results from the collisional nature of the
process. Clouds condense out of diffuse gas with a rate 
$P_{c}$ and are
disrupted by cloud-cloud collisions and by winds from massive stars, 
with rates $P_{d}$ and $P_{w}$, respectively,
\begin{equation}
\frac{{\rm d}f_c}{{\rm d}t}=P_{c} f_g^{3/2}-(P_\mathit{sf}+P_{d}) 
f_c^2-(P_{w}+P_{i}) f_sf_c,
\label{eqn:eqclouds}
\end{equation}
where $f_g(t)$ is the mass fractions of diffuse gas. 
The $3/2$ exponent results from the assumption that the process of
condensation of clouds occurs on the time scale of the gravitational
instability in the diffuse gas, proportional to the inverse square 
root of the gas density. This introduces an extra power of 1/2
in the gas mass, because mass fractions are equivalent to mass 
densities given our assumption of constant galactic volume (see below).
The mass fraction of diffuse gas evolves as
\begin{equation}
\frac{{\rm d}f_g}{{\rm d}t}=-P_{c} f_g^{3/2}+P_{d} f_c^2+ 
P_{w} f_sf_c,
\label{eqn:eqgas}
\end{equation}
where we have assumed that clouds disrupted by cloud-cloud collisions
and stellar winds produce diffuse gas. Finally, stellar remnants are
produced according to
\begin{equation}
\frac{{\rm d}f_r}{{\rm d}t}=D f_s.
\label{eqn:eqrem}
\end{equation}
Thus, the total mass of the galaxy does
not change with time ($f_g+f_c+f_s+f_r=1$, closed-box model).


\subsection{Model parameters} \label{sec:initial}

At time $t=0$ galaxies are represented as uniform-density spheres of
neutral cold gas with mass density $\rho_0$ from 
which we obtain  the initial radius of the galaxy
\begin{equation}
R_{\rm gal} = [3M_{\rm gal}/(4\pi\rho_0)]^{1/3}.
\label{eqn:rad}
\end{equation}
In the MICE models, galaxies evolve at {\em constant volume}, i.e.,
they maintain the initial radius given by Eq.~(\ref{eqn:rad}).  In
particular, we consider two values of the initial densities of atomic
gas, $\rho_0=2\times 10^{-24}$~g~cm$^{-3}$ and $\rho_0=8\times
10^{-22}$~g~cm$^{-3}$ (corresponding to numerical densities of
1~cm$^{-3}$ and 400~cm$^{-3}$) 
representing  slowly-evolving (SE) and rapidly-evolving (RE) galaxies, respectively.

We produced a grid of 30 galactic models by varying the following
two parameters:
\begin{enumerate}[a)]
\item
the initial mass: $M_{\rm gal}=10^7$, $10^8$, $10^9$, $10^{10}$, 
$10^{11}$\,\msun;
\item
the gas density in molecular clouds: 
$n_{\rm MC}=10^3$~cm$^{-3}$ (``diffuse'' clouds), 
$n_{\rm MC}=5\times 10^4$~cm$^{-3}$ (``compact'' clouds), and 
$n_{\rm MC}=10^6$~cm$^{-3}$ (``hyperdense'' clouds).
\end{enumerate} 

The molecular cloud (MC) densities are chosen to encompass the range
of volume densities observed in MC complexes
\citep[e.g.,][]{larson81,jijina99,krumholztan07}. The relatively high
densities correspond to the dense cloud cores revealed by high-density
gas tracers, since these are the true sites of star formation
\citep[e.g.,][]{lada91}.  Obviously, we cannot capture the complexity
of star formation with our models, but these densities should be
representative of the range of conditions in quiescent galaxies and
starbursts \citep[e.g.,][]{aalto95,krumholzthompson07}.  Galaxy mass is
varied to assess whether dwarf galaxies, with similar atomic and
molecular cloud properties as their more massive counterparts, can be
considered as scaled-down versions of giant galaxies, or -equivalently-
whether massive galaxies can be viewed as scaled-up versions of
dwarfs.

We associate the diffuse clouds with the passive mode of star
formation, and the hyperdense clouds with the active one.  The
distinction between active and passive modes is relatively arbitrary,
so the compact clouds defined here occupy a ``grey area'', forming
stars in an intermediate mode between ``active'' and ``passive''.
Obviously, star formation in real galaxies encompasses a wide
range of densities; these three classes are a simplification intended to
illustrate the effects of different physical conditions which could
prevail in different galaxy populations.

The  RE galaxies in our models 
correspond to the starburst mode of star formation thought to
be responsible for elliptical-galaxy assembly at
$z\simgt1-2$ \citep{delucia06,renzini06}.  
On the other hand, the SE galaxies
in our models are generally imilar to Local Universe spiral and irregular galaxies.  
In what follows, we will designate the two classes of models by
 RE galaxies and SE 
galaxies; 
these are very loosely associated with proto-ellipticals at high redshift
and spiral disks in the Local Universe. 
Our intention is to distinguish between the two ways of 
scaling the models, in the sense of spheroid proto-galaxies which
evolve rapidly and disk-like proto-galaxies which evolve more slowly.

\subsection{Scaling the models} \label{sec:scale}

The coefficients $P_{c}$, \psf, etc. in the evolution equations
Eq.~(\ref{eqn:eqstars})--(\ref{eqn:eqrem}) represent ``rates'' for the
outcomes of the various processes. Their dependence on the fundamental
parameters of the models can be estimated from educated guesses on the
dominant physical processes responsible for the conversion of gas in
clouds, stars and vice versa, as outlined below.  However, their
specific values cannot be obtained from first principles only, but are
determined in general from the application of the model to a large
sample of data.

The models adopted here have been originally developed to study the
chemical evolution of the Galaxy and nearby galaxies
\citep[e.g.,][]{ferrini92,ferrini94,molla05,magrini07,magrini09}.  Thus, the
coefficients regulating the various processes have been calibrated to
reproduce the wealth of observational data available for the Milky
Way and Local Group galaxies. In order to apply our models to the
galaxies of our sample, it is necessary to adopt simple yet robust
scaling relations for the process probabilities in terms of the 
fundamental parameters defined in Sect.\ref{sec:initial}, namely the initial 
mass $M_{\rm gal}$ and the density of molecular clouds $n_{\rm MC}$.

We start from the set of $P_{c}$ and \psf\ values adopted
to reproduce our Galaxy in \citet{magrini09}.  For the most massive
galaxy of our set (10$^{11}$\,\msun) of SE galaxies, we
adopt the $P_{c}$ and \psf\ values of the MW, while for the
RE galaxies of the same mass we increase the $P_{c}$ value
by a factor 20.  The factor 20 arises from the higher
density of the gas from which the proto-galaxy is formed, resulting in
a faster conversion of atomic gas into clouds and stars, proportional
to the inverse of the free-fall time, i.e., $\propto n_{\rm
HI}^{1/2}$.  For both classes of galaxies, in case of diffuse
molecular clouds we normalize  the value of \psf\ to the efficiency
value adopted for the MW. We thus obtain the values of $P_{c}$ and
\psf\ values for the lower mass galaxies and for galaxies hosting
denser molecular clouds with the relations described below.


In addition, we must take into account the well known result
\citep[e.g.,][]{matteucci94,calura09,pipino09} that the star formation
efficiency increases with galaxy galactic mass $M_{\rm gal}$.  
In our models, the {\em SFR efficiency} is the inverse of the depletion
timescale of the cloud mass fraction \citep[see also][]{calura09}.
This mass dependence is implied, for example, by the increasing trend of 
[Mg/Fe] with $M_{\rm gal}$ in  ellipticals \citep[e.g.,][]{nelan05} and 
from the faster evolution of large disks with respect to smaller ones
\citep{boissier01}.  This relation has taken the name of {\em chemical
downsizing} \citep{matteucci94}. It is also one of the concurring
arguments to explain the mass-metallicity relation \citep{calura09}.
\citet{calura06} suggested that the main features of local galaxies of
different morphological types may be reproduced by interpreting the
Hubble sequence as a sequence of decreasing star formation efficiency
from early types to late types, i.e. from ellipticals to irregulars
\citep[see also][for an early interpretation of the Hubble sequence in
terms of SFRs]{galli89}.  

In our models we have scaled the rate of cloud condensation as
$P_c\propto M_{\rm gal}^{1/3}$ and the rate of star formation as
\psf$\propto M_{\rm gal}^{1/3} n_{\rm MC}^{1/2}$.  The scaling with
mass corresponds basically to a scaling with size, and, at least for the
parameter \psf, follows naturally from the interpretation of this
quantity as a {\it collision frequency} between clouds. The latter is
given by the product of the number of clouds per unit volume (roughly
proportional to $M_{\rm gal}\,R_{\rm gal}^{-3}$) times the cloud's cross
section (assumed the same in all models) and the cloud-to-cloud
velocity dispersion (proportional to $M_{\rm gal}^{1/2}\,R_{\rm
gal}^{-1/2}$).  However, the SFR efficiency also depends 
on the dynamical (free-fall) time of the molecular clouds 
($\propto n_{\rm MC}^{-1/2}$), which results in values of \psf\
larger by a factor of $\sim 7$ and $\sim 32$ in the case of 
compact and hyperdense clouds, respectively, with respect
to diffuse clouds.  
Observing that what we call
``diffuse'' clouds are typical of the MW and nearby quiescent disk
galaxies \citep{simon01,krumholzthompson07}, we can just scale the
value of \psf\ adopted for the MW (taken as a reference) to
obtain the \psf\ parameter in the other two cases.  This
scaling approximately reproduces the observed fraction of gas
($\sim$1\%) that forms stars during a free-fall time
\citep{krumholztan07}.

In summary, our models consist of two classes of galaxies:
slowly- and rapidly-evolving galaxies.
For SE galaxies, we normalized the
$P_{c}$ coefficient of the most massive ones ($M_{\rm
gal}=10^{11}$\,\msun) to the coefficient of the MW model of
\citet{magrini09}; for RE galaxies, we adopted the
same coefficient multiplied by a factor of 20.  For both classes of
galaxies, in the case of diffuse molecular clouds we normalized  the
\psf\ to the value adopted for the MW. We then derived the $P_{c}$ and
\psf\ coefficients for the lower mass galaxies and for galaxies hosting
denser molecular clouds by adopting the scaling \psf $\propto M_{\rm
gal}^{1/3}n_{\rm MC}^{1/2}$ and $P_{c}\propto M_{\rm gal}^{1/3}$.  We
emphasize, however, that our scalings are largely empirical, and must
be considered as a rough parametric representation of physical
processes not included in detail in our model (internal dynamics, mass
loss and accretion, etc.).  In Table~\ref{tab:par} we present a summary
of our model scaling factors with respect to the MW values (P$_{c,{\rm MW}}$
and P$_{\mathit{sf},{\rm MW}}$) for each type of galaxy. 


\begin{table}
\caption{Model scaling factors with respect to the MW values.}
\begin{tabular}{llllll}
\hline
\multicolumn{5}{c}{}\\
M$_{\rm gal}$		& $P_{c}/P_{c,{\rm MW}}$ & $P_{\mathit sf}/P_{\mathit{sf},{\rm MW}}$ & ${P_{\mathit sf}}/P_{\mathit{sf},{\rm MW}}$ &$P_{\mathit sf}/P_{\mathit{sf},{\rm MW}}$    \\  
(M$_{\odot}$)	         &                        &   Diffuse                     & Compact                        &Hyper-dense\\	
\multicolumn{5}{c}{}\\
\hline
\multicolumn{5}{c}{Rapidly-evolving galaxies ($n_{\rm HI}=400$~cm$^{-3}$)}\\
10$^{11}$		& 		  20.   & 1           & 7    & 32\\
10$^{10}$		&		  9.6   & 0.46     & 3.2 & 15\\
10$^{9}$		&		  4.4   & 0.22     & 1.5 & 7\\
10$^{8}$		&		  2.0   & 0.10     & 0.7 & 3.2\\
10$^{7}$		&		  0.8   & 0.02     & 0.1 &0.6\\
\hline
\multicolumn{5}{c}{Slowly-evolving galaxies ($n_{\rm HI}=1$~cm$^{-3}$)}\\
10$^{11}$		& 		  1.   & 1           & 7    & 32\\
10$^{10}$		&		  0.46   & 0.46     & 3.2 & 15\\
10$^{9}$		&		  0.22   & 0.22     & 1.5 & 7\\
10$^{8}$		&		  0.10  & 0.10     & 0.7 & 3.2\\
10$^{7}$		&		  0.02   & 0.02     & 0.1 &0.6\\
\hline
\hline
\multicolumn{5}{c}{}\\
\end{tabular}
\label{tab:par}
\\
\end{table}

\subsection{Stellar yields and IMF}

The chemical enrichment of the gas is modeled using the matrix
formalism developed by \citet{talbot73}. The elements of the
restitution matrices $Q_{i,j}(M,Z)$ are defined as the fraction of the
mass of an element $j$ initially present in a star of mass $M$ and
metallicity $Z$ that it is converted into an element $i$ and ejected.
This version of the MICE model takes into account two different
metallicities, $Z=0.02$ and $Z=0.006$, and 22 stellar mass bins (21 for
$Z=0.006$) from $M_{\rm min}=0.8$\,\msun\ to $M_{\rm
max}=100~$\,\msun, for a total of 43 restitution matrices.  For low-
and intermediate-mass stars ($M<8$\,\msun) we use the yields by
\citet{gavilan05} for both values of the metallicity. For stars in the
mass range $13~M_\odot < M < 35~M_\odot$ we adopt the yields by
\citet{chieffi04} for $Z=0.006$ and $Z=0.02$.  We estimate the yields
of stars in the mass range $8~M_\odot < M < 13~M_\odot$ 
$35~M_\odot <M<100~M_\odot$, which are not
included in tables of \citet{chieffi04} by linear extrapolation of the
yields in the mass range $13~M_\odot < M < 35~M_\odot$.
The SNIa yields are taken from the model CDD1 by \citet{iwamoto99}.

The choice of different sets of stellar yields affects the final results 
as highlighted by \citet{romano10}. 
In the present work, we are mainly interested on the evolution of the oxygen abundance, 
whose nucleosynthesis in stars is relatively well understood. 
Thus, with different prescriptions for the O production we  
do not obtain dramatically different results in terms of O/H, although, 
as demonstrated  by \citet{romano10},
the [O/Fe] ratio may vary 0.2\,dex at solar metallicity, and even more at 
lower metallicities. 

Another fundamental ingredient is the IMF.
Several parametrizations of the IMF have been widely used in the
literature.  In the present version of the code we use the
\citet{chabrier03} mass function in the mass range 0.1-100\,\msun.

\section{Comparison of model predictions with observations} \label{sec:comparison}

We can now address the {\em active} and {\em passive} SF modes 
of galaxies with similar masses and metallicities. 
We first illustrate the time evolution of the stellar
mass \mstar, the SFR, and the metallicity \logoh.
We then discuss the models for the observed scaling relations described in Paper~I,
namely the SFMS, or correlation between \mstar\ and sSFR,
and the MZR, or mass-metallicity relation.

\subsection{The time evolution of O/H, SFR, and \mstar}

\begin{figure*}
\hbox{
\includegraphics[width=0.33\linewidth,bb=18 144 592 650]{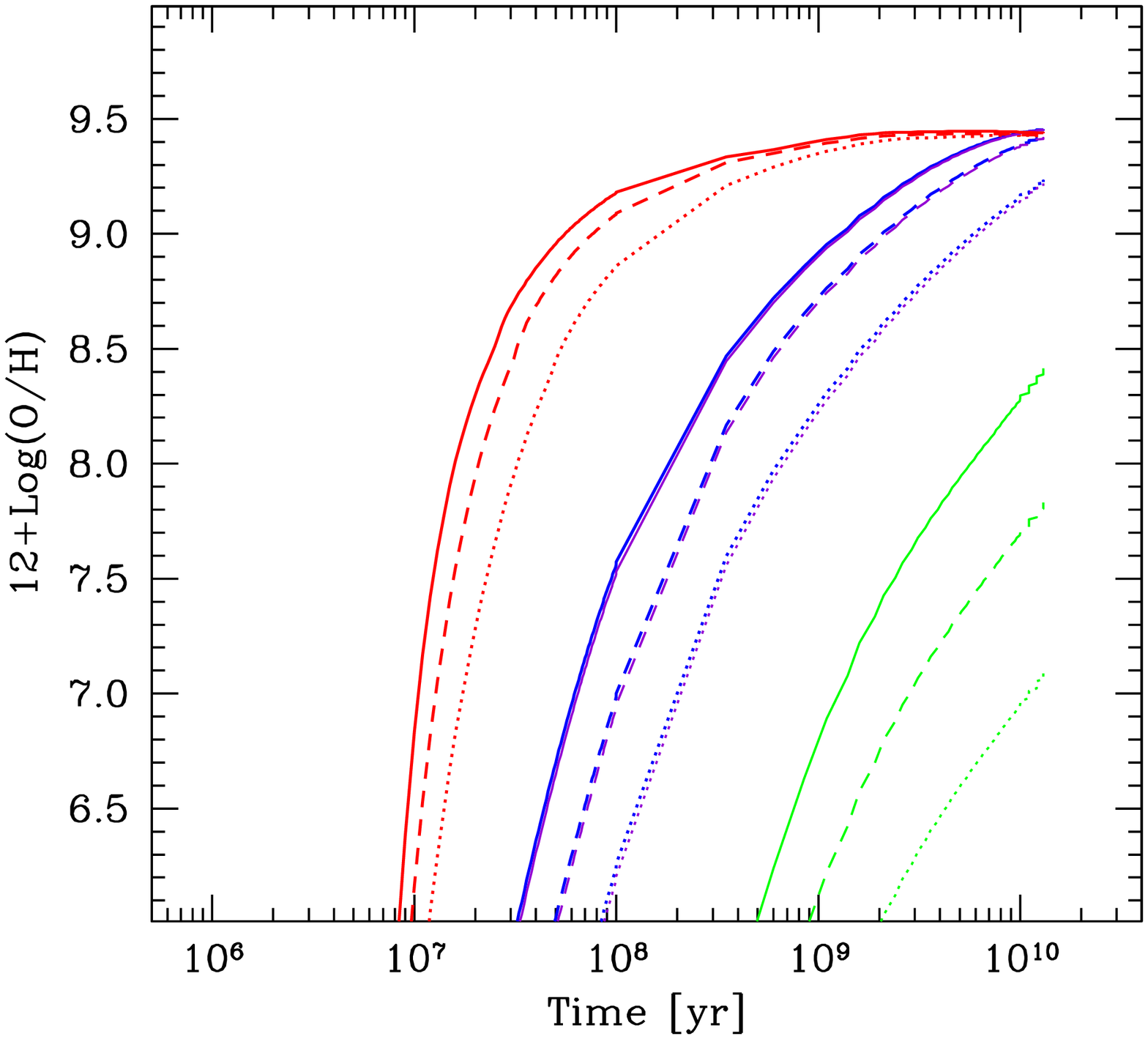} 
\includegraphics[width=0.33\linewidth,bb=18 144 592 650]{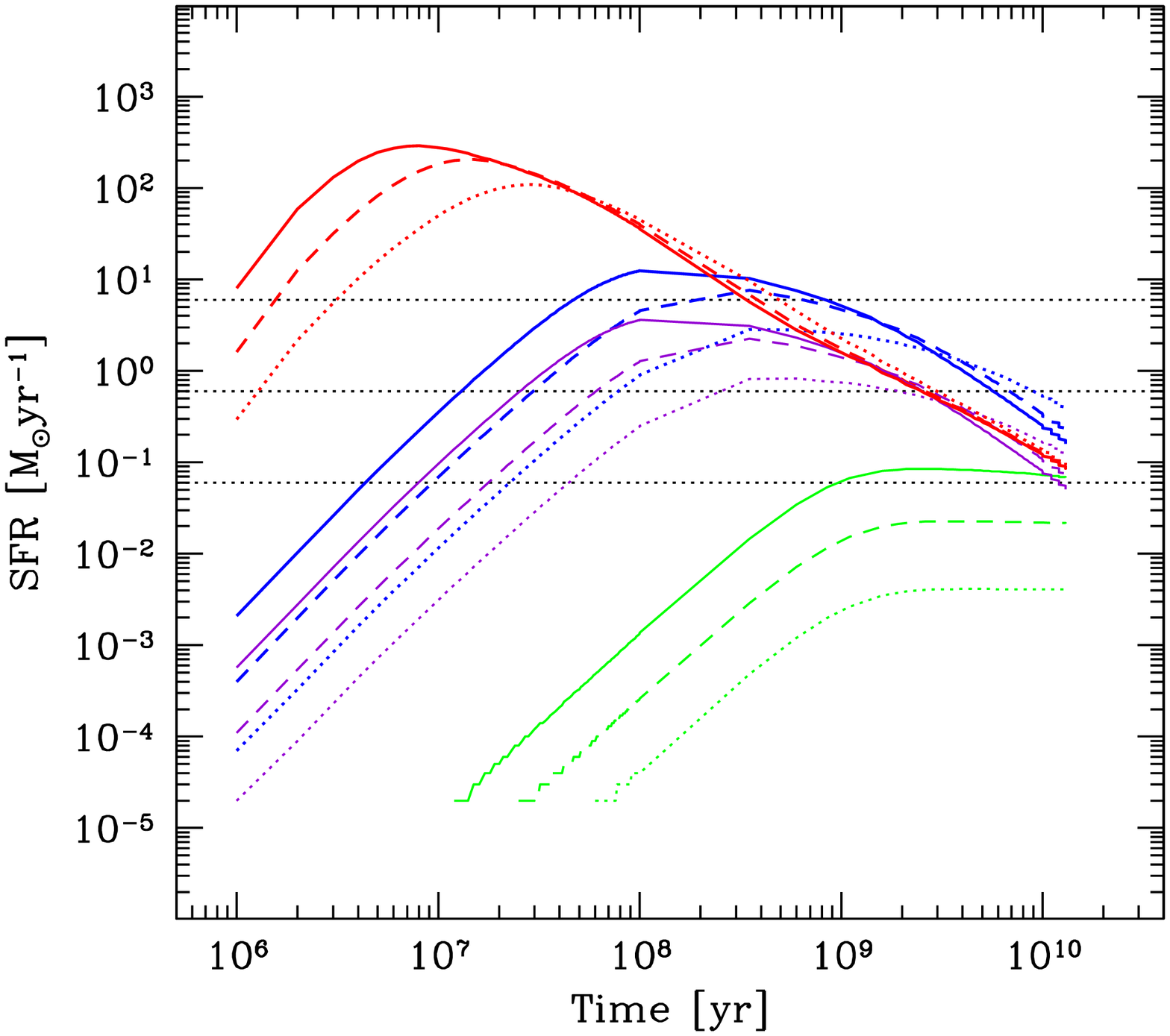} 
\includegraphics[width=0.33\linewidth,bb=18 144 592 650]{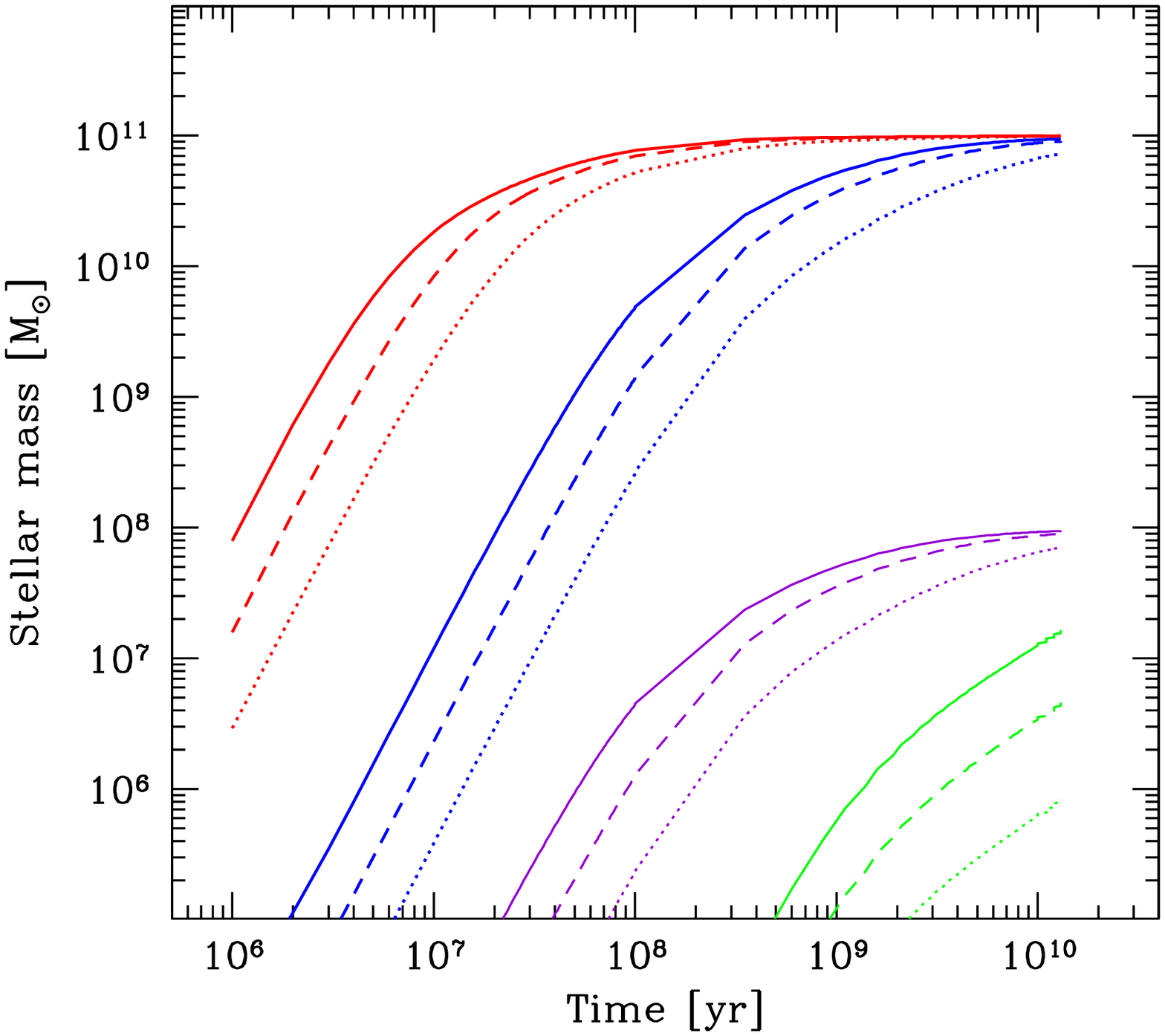} 
}
\caption{Predictions of O/H (left panel), SFR (middle panel), and \mstar\ (right
panel) vs. time in the MICE models.
Models of two representative galaxies, $10^8$\,\msun\ and $10^{11}$\,\msun, 
are coded by line type and color.  
The diffuse, compact, and hyper-dense cases are shown
as dotted, dashed, and solid lines, respectively. 
Different galaxy masses are distinguished by color:
green/dark violet for $10^8$\,\msun,
and blue/red for $10^{11}$\,\msun\ (SE/RE, respectively).
In the middle panel, dotted horizontal lines show the four SFR regimes considered
here: $\mbox{SFR}\leq 0.06$\,\msunyr, $0.06 < \mbox{SFR}\leq 0.6$\,\msunyr, 
$0.6 < \mbox{SFR} \leq 6$\,\msunyr, and $\mbox{SFR}\geq 6$\,\msunyr.
}
\label{fig:sfh_time}
\end{figure*}

The time evolution of  metallicity, SFR and \mstar\ are shown  in
Fig.~\ref{fig:sfh_time} for two
sets of galactic models with different initial masses  
$M_{\rm gal}=10^{8}$ (shown as dark violet/green, for SE/RE, respectively)
and $10^{11}$\,\msun\ (blue/red, SE/RE).  For each mass, we show 
MCs with three different densities: diffuse 
(shown as dotted lines), compact (dashed lines), and 
hyperdense (solid lines). 

For a given galactic mass, $M_{\rm gal}$, 
RE galaxies evolve much more rapidly than SE galaxies. 
They accumulate metals and
build a considerable fraction of their stellar mass\footnote{The 
stellar mass plotted here is the ``luminous mass'', and does not take into
account the mass of the stellar remnants, white dwarfs and neutron
stars.} 
in a shorter time ($\sim
10^8$~yr for $M_{\rm gal}=10^{11}$\,\msun, and $\sim 10^9$~yr for
$M_{\rm gal}=10^8$\,\msun). Slowly-evolving disk-like galaxies of the same mass need
much more time for metal production and to assemble stellar mass
($\sim 10^9$~yr and more than a Hubble time to produce the same stellar
fraction, respectively).  
This signature of {\em chemical downsizing}, here closely
related to mass downsizing, stems from the 
considerable fraction of the stellar mass in massive galaxies formed in less 
than 1\,Gyr, while 50\% or more of the stellar mass of less
massive galaxies is still being created at the present time.
This is because of the high efficiency of star formation
of  RE galaxies in their formation phase.  

For a given galactic mass, 
SFR also peaks at earlier times for RE galaxies  with respect to SE galaxies,
as already found by \citet{sandage86}. 
Galaxies with denser molecular clouds (i.e., a more active SF mode) have a
higher maximum value of the SFR with an earlier onset than galaxies with less dense clouds. 
Within either SE or RE galaxies, for a specific cloud density, 
more massive galaxies reach a higher SFR and at earlier times than less massive galaxies
\citep[because of their increased star-formation efficiency, see also][]{tosi85}. 

At a given molecular cloud gas density, 
$n_{\rm MC}$, more massive galaxies evolve more rapidly
than less massive ones (e.g., {\em downsizing}\/).  
For the most massive galaxies, 
$M_{\rm gal}=10^{11}$\,\msun, the final oxygen abundances are similar for
the rapidly- and SE families, with slightly lower values for
 SE galaxies, even though these values are reached over different
timescales.  For the least massive galaxies with $M_{\rm gal}=10^8$\msun,
SE  and RE galaxies achieve different metallicities at the present
epoch.  Again, this is because of chemical downsizing, since
less massive galaxies form stars less efficiently. Thus, although
SE disk-like galaxies with $M_{\rm gal}=10^8$\msun\ have the potential to
reach the same metallicity of RE galaxies of the same mass, 
they will do so only on timescales much longer than a Hubble time.
They also require more than a Hubble time to fully convert all their gas
into stars.

Within each class (SE, RE), galaxies with the densest
clouds evolve faster.
These differences again result from the star formation
efficiency which, in our models, depends also on the free-fall time of the
molecular clouds.  

The left panel of Fig.~\ref{fig:sfh_time} shows the degeneracy
between the track of the  SE galaxy with $M_{\rm gal}=10^{11}$\,\msun\
and that of the RE galaxy with $M_{\rm gal}=10^8$\,\msun.  The
scaling factors described in Sect.~\ref{sec:scale} conspire to make the
behaviour of these two galaxies virtually identical in terms of 
metallicity evolution. However, the evolutionary timescales
of the SFR and the stellar mass differ, 
suggesting that metallicity alone is not a good indicator of
the evolutionary status of a galaxy. Specifically, a slowly evolving
massive galaxy and a low-mass rapidly evolving galaxy will have similar metallicities. 

Indeed, Fig.~\ref{fig:sfh_time} also shows that for a given galactic mass, it
is possible to achieve the same metal abundance (left panel)
with different SFRs (middle) and different stellar masses (right).
The galaxy can be less massive but, because of its active mode,
more rapidly form stars, and thus accumulate more metals and stellar mass on 
shorter timescales. Alternatively it can be more massive 
and thus produce more of everything, but can also form stars and accumulate 
metals more slowly.
The key aspect of our modeling is the {\em evolutionary timescale} 
which is parametrized either through the global properties
of the galaxies, governed by how we have scaled our models
(i.e., slow vs. rapid evolution), 
or through the density of the gas in the molecular clouds within the galaxies.
Either way, shorter dynamical times drive more rapid evolution. 
We will discuss below how these effects
can help explain the {\em active} vs. {\em passive} dichotomy,
and consequently how we can better understand the behavior of low-metallicity 
starbursts, and their departure from the scaling relations.

In any case, the present-day abundances of our model galaxies are somewhat high.
This is almost certainly due to the closed-box nature of our models
and the lack of gas exchange with the environment.
In fact,
this is what led chemical evolution modelers to introduce both infall of metal poor
gas and later also galactic winds powered by supernova explosions
in order to reduce the predicted metal
abundances and bring them to a better agreement with the observed ones
\citep{matteucci85,tosi88,pilyugin93}.
Without galactic winds the products of stellar nucleosynthesis are always
retained, and without infall of pristine gas they are never diluted.
It is well known that gas accretion and expulsion can significantly alter 
the chemical abundances in galaxies
\citep[e.g.,][]{matteucci85,pilyugin93,marconi94,romano06}.

However, infall and/or outflow are not the only mechanisms  
that could vary the abundances. The absolute level of the abundance depends also  
on the combination of the adopted stellar yields and IMF.
\citet{magrini10} have shown that the maximum variation of the  
abundance level in the disk of the
spiral galaxy M\,33 obtained by varying the parametrization of the IMF (and  
keeping fixed the stellar yields) is $\sim$0.6\,dex,
with the Chabrier IMF having the higher abundances. The impact of  
adopting different stellar yields, e.g.,
considering or not stellar rotation, could be $\sim$0.2\,dex at  
solar metallicity \citep[c.f.,][]{romano10}.
Thus, considering quadratically the two effects, 
the absolute abundance level could vary by up to $\sim$0.6\,dex
without invoking infall/outflow processes.
In any case,
the approach presented here should be considered as a first step toward
better understanding the interplay of the three observables; future work
will be aimed at comparing our current results
with models having gas inflow and outflow and with different IMFs and stellar yields.

\begin{figure*}
\hbox{
\includegraphics[width=\linewidth,bb=18 144 592 718]{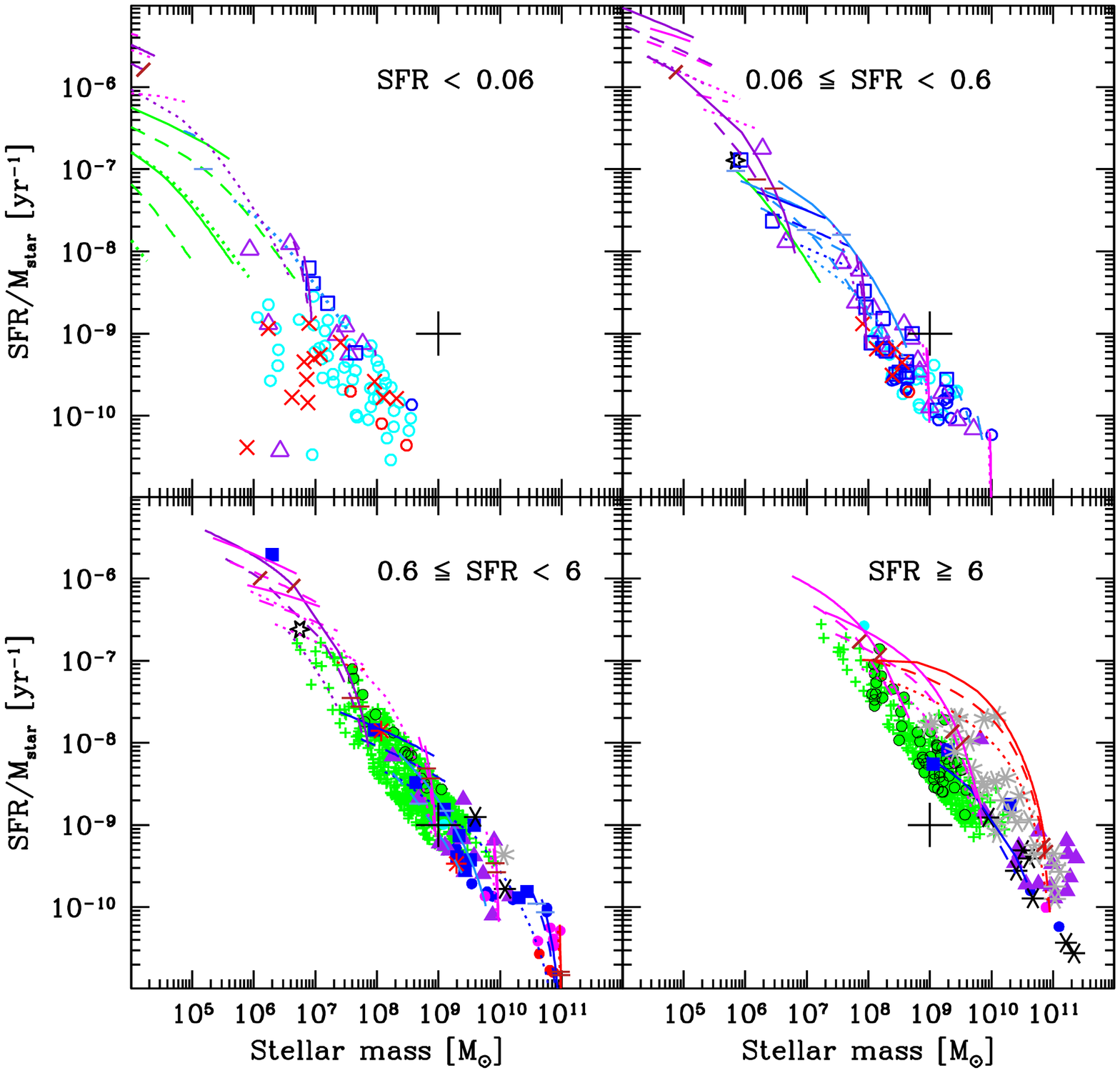} 
}
\caption{Specific SFR vs. stellar mass divided into four ranges of
SFR:  $\mbox{SFR}\leq 0.06$\,\msunyr, $0.06 <\mbox{SFR}\leq
0.6$\,\msunyr, $0.6 < \mbox{SFR}\leq 6$\,\msunyr, $\mbox{SFR}\geq
6$\,\msunyr.  
LVL (11HUGS) galaxies are shown as small open or filled circles
(filled when SFR$\geq$0.6\,\msunyr),
with different colors corresponding to Hubble type $T$ as in
\citet{lee09}: $T \geq$ 8 cyan, $5\leq T<8$ blue,  $3\leq T<5$ magenta,
and $T<3$ red.
Red $\times$ corresponds to the dIrr sample;
BCDs are shown as (blue) squares (Fumagalli$+$Hunt samples) and 
(purple) triangles (Engelbracht);
Green Peas are given by small (green) filled circles; LCGs by (green) $+$;
LBGs at $z\sim1$ (Shapley), $z\sim2$ (Erb, Shapley) are shown as 6-pronged asterisks, 
and at $z\sim3$ (Maiolino, Mannucci) as 8-pronged asterisks. 
6-sided open stars show the two dwarf ``prototypes'', \sbs\ and \izw. 
Solid symbols show those BCDs with SFR$\geq$0.6\,\msunyr.
Models are coded by line type and color as in Fig.~\ref{fig:sfh_time}.
The corresponding colour codes for the stellar masses (SE/RE, respectively)
are: green/dark violet for $10^7$, $10^8$\,\msun,
light blue/magenta for $10^9$, $10^{10}$\,\msun,
and blue/red for  $10^{11}$\,\msun.
The evolution times are marked along the tracks: 
an inclined tick indicates 100\,Myr (only for the RE galaxies), 
and a horizontal tick 1\,Gyr. 
Time increases downward and to the right.
The large $+$ sign is a fiducial point (sSFR = 10$^{-9}$\,yr$^{-1}$,
\mstar\,=\,10$^9$\,\msun) which guides the eye to the differences
in the main loci of data points and models in the different panels.
} 
\label{fig:ssfr_mass_4p} 
\end{figure*}

\subsection{The SFMS and the MZR }

Here we compare the model results with the observed
SFMS (sSFR vs \mstar\ relation) and the MZR (mass-metallicity relation). 
We also show the models superimposed on the Fundamental Plane identified
in Paper~I.
To better visualize the interplay among the observables,
and isolate different regions of parameter space,
we have divided the models and the data into two regimes of SFR:
the ``passive'' one with SFR$<$0.6\,\msunyr, and the ``active'' one
with SFR$\geq$0.6\,\msunyr.
This threshold was defined in Paper~I on the basis of the onset of deviations 
from the SFMS and the MZR; for our sample, these deviations become significant
at all redshifts for SFR$\ga$0.6\,\msunyr, roughly the equivalent of
NGC\,4826 (M\,64, the Black-Eye Galaxy), 1.5 times M\,33 or 
the ``dwarf starburst'' NGC\,3077, or twice the SFR of M\,31 
\citep{dale12,verley07,meier01,tabata10}.
We have further divided these two regimes into two sectors, in order
to better separate and identify the progression from passive to active modes.
In all of the following plots, we show models at times from 1\,Myr to a Hubble time,
as given in Fig.~\ref{fig:sfh_time}. 


\subsubsection{The main sequence of star formation}

Figure~\ref{fig:ssfr_mass_4p} shows the SFMS with sSFR plotted against
stellar mass.
Galaxies characterized by a {\em passive} SF mode appear
in the upper panels, while {\em active} galaxies are in the
bottom ones. 
Models of RE and SE galaxies, with
their mass ``families'' and three different MC density regimes, are also
shown and divided into different SFR regimes. 
Evolution timescales are marked along the tracks: 
an inclined tick indicates 100\,Myr (only for the RE galaxies), 
and a horizontal tick 1\,Gyr. 
The $+$ marks an arbitrarily defined fiducial position in order
to guide the eye to the displacements in the various panels as
a function of SFR (see below).

The least massive most quiescent galaxies (see upper left panel)
have very low sSFRs, and are modeled only by the very end points
of SE galaxies.
As SFR increases (upper right), 
galaxies are more massive and the data are well reproduced by  
SE models with either low masses and compact MCs (the ``intermediate'' SF mode), 
or higher masses and diffuse clouds (the passive mode).
Data in the bottom left panel (the onset of the {\em active} SF mode), 
are well modeled by intermediate-mass SE  or RE
galaxies
with dense or hyper-dense clouds, but also by the most massive SE galaxies
(or RE galaxies) with diffuse clouds.
As discussed in Sect.~\ref{sec:scale}, both galaxy mass and cloud density
contribute to SFR and evolutionary timescales.

The most extreme starbursts in the bottom right panel
are reproduced only by RE galaxies and the most
massive SE galaxies with hyper-dense and compact clouds. 
Moreover, such high SFRs
are achieved by the most massive RE models
and the most massive hyper-dense  SE only up to an
age of $\simlt$200--500\,Myr from the onset of the SF episode
(see Fig.~\ref{fig:sfh_time}). 
Thus galaxies with these high SFRs are expected to be ``young'', 
i.e., ``captured'' in a relatively early phase of their evolution. 
However, the RE models 
have assembled almost all their stellar mass (and metals)
already at these early times, but the SE models will continue 
to build up their stellar and metal content for some time to come.

The $+$ in Fig.~\ref{fig:ssfr_mass_4p} represents a 
fiducial point (arbitrarily defined), 
with the aim of guiding the eye to the changes in loci among the panels.
In the upper left panel, the galaxies fall below the fiducial by
almost an order of magnitude in both axes; in the upper right and
lower left, the galaxies (and models) are roughly coincident; in the
last panel both galaxies and models fall above the fiducial by a factor of
5 or so in both sSFR and stellar mass.
Many LCGs and Green Peas at $z\sim0.3$, and some BCDs at $z\sim0$, 
are coincident with the LBGs in the AMAZE and LSD samples at $z\sim3$.
These galaxies are modeled by systems in early stages of their
evolution, and with high SFR. Hence, at least in this parameter space
of SFR and \mstar, there are local galaxy populations which mimic the properties
of galaxies at high redshift.
They are distinguished by their relatively high SFR, perhaps resulting from a 
similar selection effect for both the low- and high-redshift samples.

\subsubsection{The mass-metallicity relation}

\begin{figure*}
\hbox{
\includegraphics[width=\linewidth,bb=18 144 592 718]{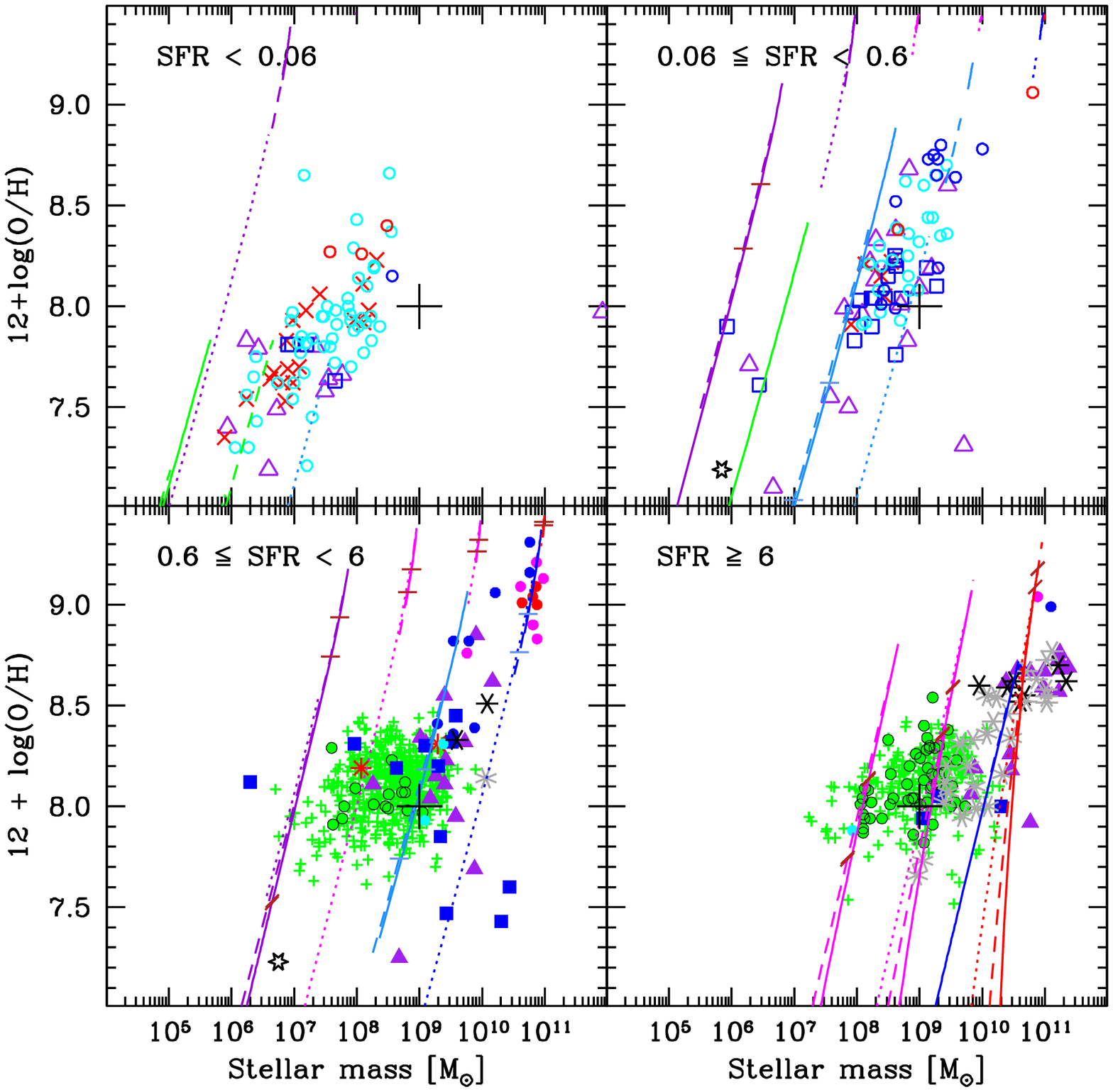} 
}
\caption{Nebular oxygen abundance vs. stellar mass $M_{\rm star}$
divided into regions of SFR.  The four regimes we consider are:
$\mbox{SFR}\leq 0.06$\,\msunyr, $0.06<\mbox{SFR}\leq 0.6$\,\msunyr, $0.6
<\mbox{SFR}\leq 6$\,\msunyr, $\mbox{SFR} \geq 6$\,\msunyr.  
Symbols are as in Fig.~\ref{fig:ssfr_mass_4p}.
Models are coded by line type and color as in Fig.~\ref{fig:ssfr_mass_4p}:
diffuse as dotted lines, compact as dashed, and hyper-dense as solid lines.
The corresponding colour codes for the stellar masses (SE/RE, respectively)
are: green/dark violet for $10^7$, $10^8$\,\msun,
light blue/magenta for $10^9$, $10^{10}$\,\msun,
and blue/red for  $10^{11}$\,\msun.
The evolution times are marked along the tracks: 
an inclined tick indicates 100\,Myr (only for RE galaxies), 
and a horizontal tick 1\,Gyr. 
Time increases upward and to the right.
The large $+$ corresponds to a ``fiducial'' value of \mstar\,=\,10$^9$\,\msun,
and \logoh\,=\,8.0, intended to guide the eye to differences among
the panels as in Fig.~\ref{fig:ssfr_mass_4p}.
}
\label{fig:oh_mass_4p}
\end{figure*}

\begin{figure*}
\hbox{
\includegraphics[width=\linewidth,bb=18 144 592 600]{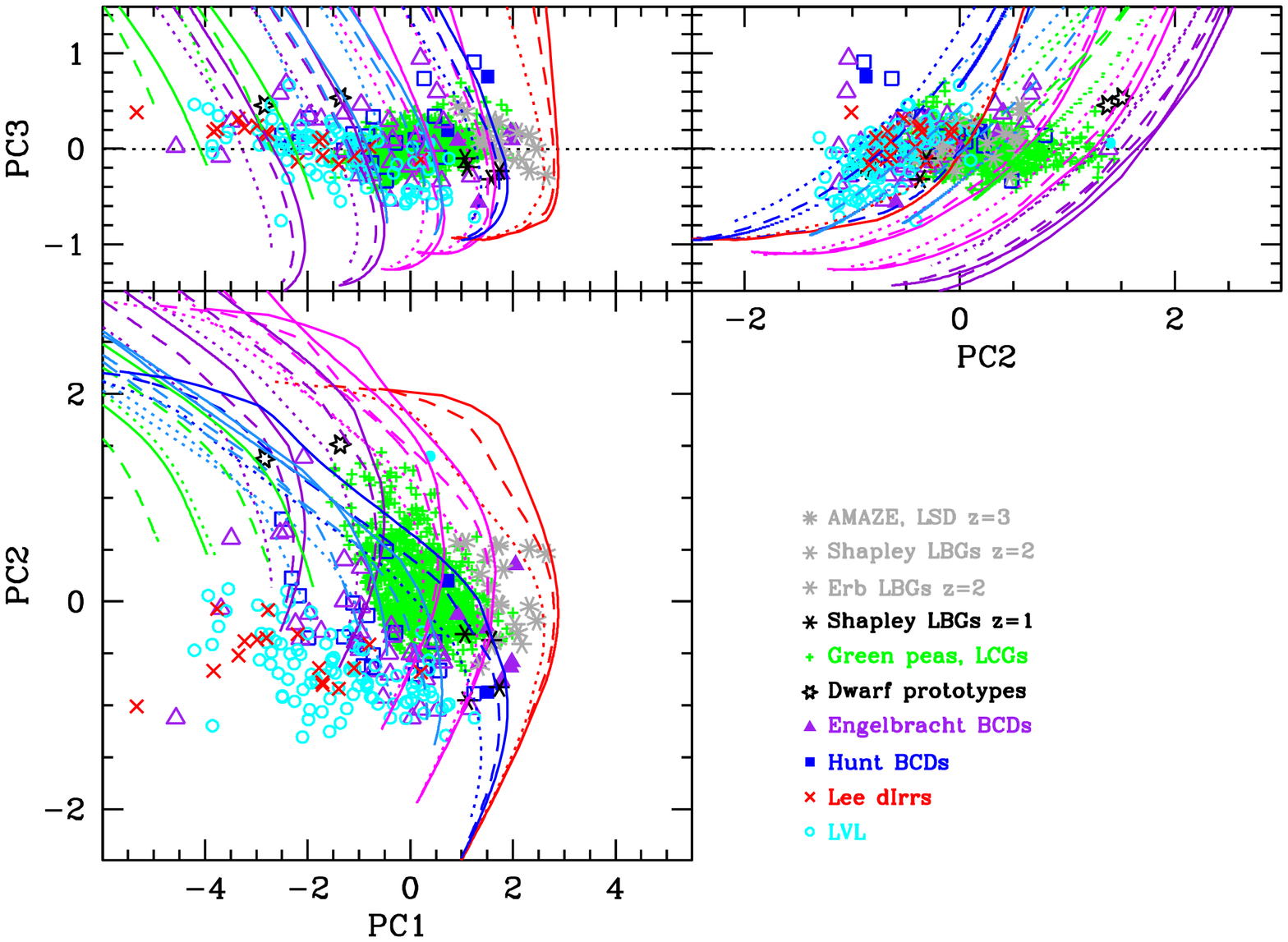} 
}
\caption{Different projections using the 3 PCs found by the PCA.
Galaxies are coded as shown in the lower right (empty) panel.
The top left and right panels show the orthogonal ``edge-on'' views of the plane; 
the bottom panel shows the plane face-on.
$x_1\,=\,$\logoh$-\langle$ \logoh $\rangle$,
$x_2\,=\,\log (SFR)-\langle \log(SFR) \rangle$ (\msunyr), and
$x_3\,=\,\log (M_{\rm star})-\langle \log(M_{\rm star}) \rangle$ (\msun).
$\langle$\logoh$\rangle$\,=\, 8.063;
$\langle\log(SFR)\rangle$\,=\, -0.594;
$\langle\log(M_{\rm star})\rangle$\,=\,8.476.
PC1 \, = \, 0.12\,$x_1$ + 0.75\,$x_2$ + 0.65\,$x_3$; 
PC2 \, = \, -0.31\,$x_1$ - 0.65\,$x_2$ - 0.69\,$x_3$;
PC3 \, = \, -0.94\,$x_1$ - 0.11\,$x_2$ + 0.31\,$x_3$.
The axes are expanded in the top panels to exaggerate the variations in PC3,
relative to the much larger dynamic ranges in PC1 and PC2.
MICE models are superimposed similarly to previous figures, but here
there are no temporal tick marks. 
In all three panels, time increases downward.
}
\label{fig:pca}
\end{figure*}

The MZR is shown in Fig.~\ref{fig:oh_mass_4p}, divided into four SFR regimes 
as in Fig.~\ref{fig:ssfr_mass_4p}.
Low-mass galaxies with relatively high metallicities (top left panel) 
are not well reproduced with our MICE models.
This is probably because our SE models require more than
a Hubble time to complete the stellar mass assembly and metal production
with the available fuel. 
Ongoing accretion could help here, but it would also be necessary to shorten 
timescales in order to accelerate the gas consumption.
As SFR increases (top right), intermediate-mass SE  galaxies are able to 
well reproduce the data, either more massive models with diffuse clouds 
or less massive ones with compact or hyper-dense clouds.

Galaxies forming stars in the active mode (bottom left panel) are well reproduced
by the more massive SE  galaxies and the less massive RE galaxies.
The most massive SE  galaxy  (10$^{11}$\,\msun) with diffuse clouds is consistent with
the most metal-rich observations, and also the most metal-poor ones.
Interestingly,
the latter appear to lie along the same evolutionary path as the former 
and may evolve into them.
However, an active SF mode with compact or hyper-dense clouds
is necessary to reproduce the observations at lower
masses: either  SE  galaxies for masses $10^8-10^9$\,\msun, or
RE galaxies for lower masses.

Extreme starbursts (bottom right panel) require rapidly evolving models 
at young ages for such high SFRs; galaxies must be younger than $\simlt$200-300\,Myr 
(see Fig.~\ref{fig:sfh_time}). 
The main population of the
galaxies in this panel can be reproduced only 
with the early phases of the evolutionary tracks of 
RE (starburst) galaxies. 
Only a single SE galaxy model, 
the most massive one with hyper-dense molecular clouds, 
appears in this panel.
Below,
we explore how the gas content of the models changes with SFR (and age)
in order to better understand the evolutionary phase of these galaxies.

\subsection{The fundamental plane of metallicity, SFR, and stellar mass}\label{sec:fp}

Our models have been rather successful at reproducing the observed trends in
the SFMS and MZR.
However, in Paper~I, we introduced a Fundamental Plane (FP) which enabled an accurate
(0.17\,dex) determination of O/H given SFR and \mstar.
Here we show that our models also well reproduce the FP, and 
occupy the same spread in parameter space as the observations.

Figure~\ref{fig:pca} shows the Fundamental Plane that
emerged from the Principle Component Analysis (PCA) of Paper~I.
The two upper panels show the edge-on views of the plane, and the bottom
panel the face-on one.
Our models are superimposed on the data and are consistent with the loci
of the data in all three projections.
In all three plots, time increases downward. 

\section{Gas fractions and metal abundance}\label{sec:gas}

\begin{figure*}
\hbox{
\includegraphics[width=0.5\linewidth,bb=18 144 592 650]{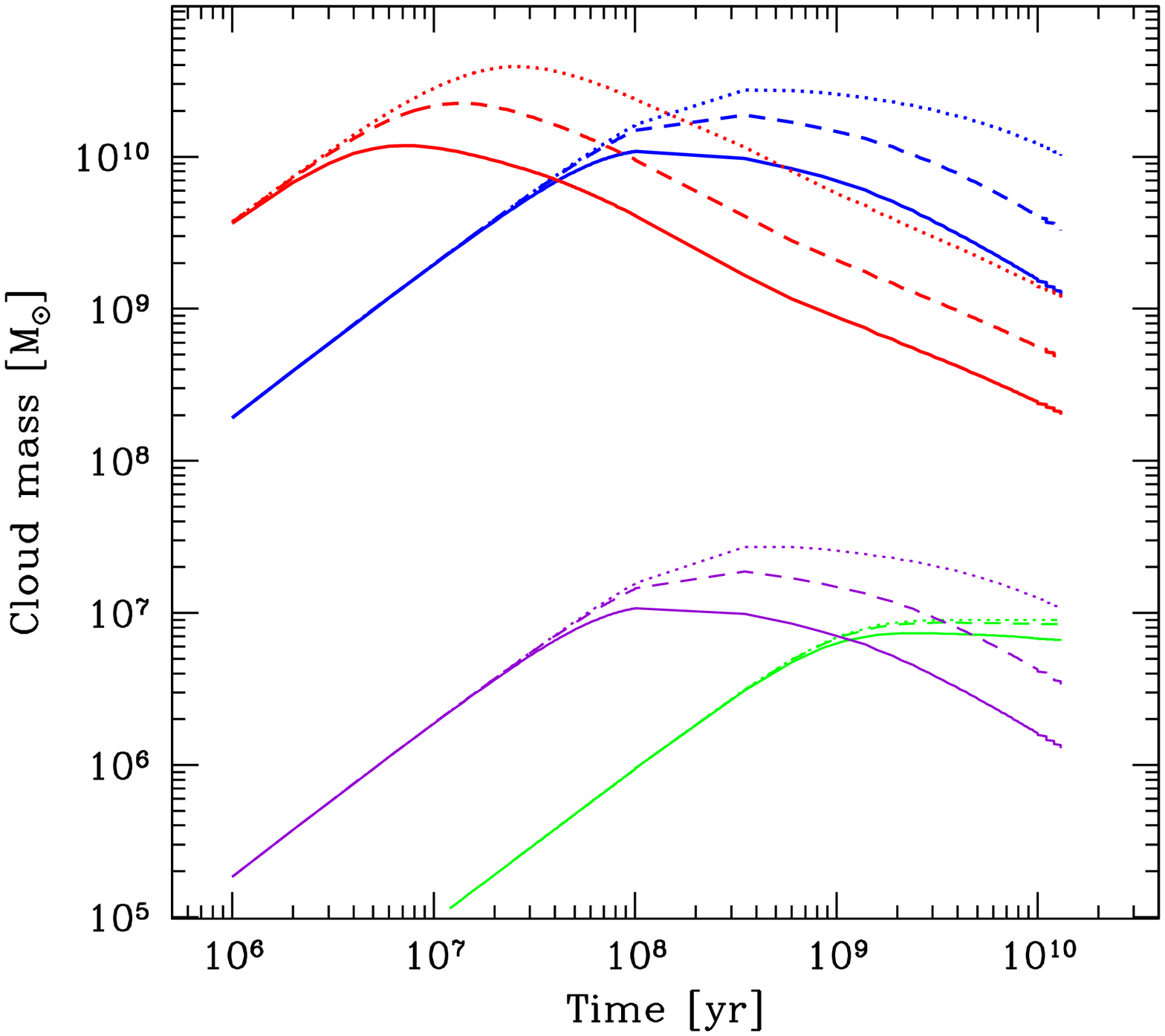} 
\includegraphics[width=0.5\linewidth,bb=18 144 592 650]{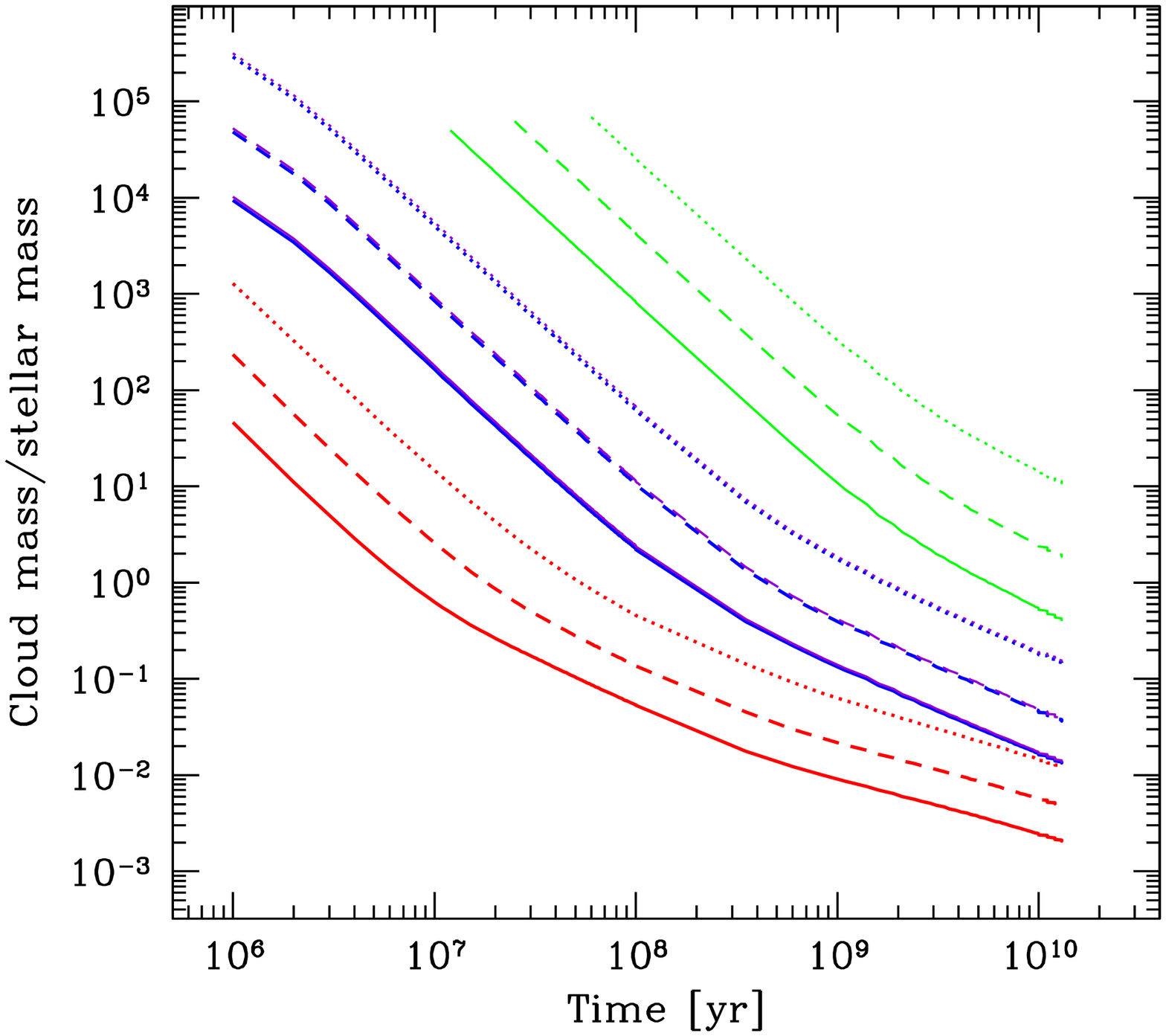} 
}
\caption{Predictions of \htwo\ mass (left panel) and ratio of \htwo\ to \mstar\ (right
panel) vs. time in the MICE models.
As in Fig.~\ref{fig:sfh_time}, we show only
models of two representative galaxies, $10^8$\,\msun\ (green/dark violet)
and $10^{11}$\,\msun\ (blue/red), 
for the SE/RE types, respectively.  
The diffuse (dotted lines), compact (dashed), and
hyper-dense cases (solid lines) are coded by line type.
In these models, the cloud-formation efficiency and SFR efficiency
both scale with cloud mass as: $M_{\rm MC}^{1/3}$.
}
\label{fig:gas_time}
\end{figure*}

Most previous interpretations of the shape of the
MZR have relied on gas inflow and outflow, with the assumptions
that outflowing gas is chemically enriched and that
infalling gas is not
\citep[e.g.,][]{tolstoy09,mannucci10,dave12,yates12,dayal12}.
As already discussed,
our MICE models are ``closed-box'', and thus neglect the effects of gas inflow and outflow.
Nevertheless, as we have shown in previous sections,
the predictions of our models are in good agreement with the SFMS and the MZR
through the active/passive dichotomy of SF modes and the resulting evolutionary
timescales.
Here we examine the MICE model predictions for
the molecular gas mass fractions relative to stellar mass,
and compare them with observations of local and high-$z$
samples with measured \htwo\ and stellar masses.

The time evolution of the \htwo\ content (``cloud'' mass) and the 
cloud-to-stellar mass ratio is shown in Fig.~\ref{fig:gas_time}.
The right panel of this figure, as in Fig.~\ref{fig:sfh_time},
shows the degeneracy
between the track of the SE  galaxy with $M_{\rm gal}=10^{11}$\,\msun\
and that of the RE galaxy with $M_{\rm gal}=10^8$\,\msun.  
The scaling factors conspire to make these two galaxies behave
similarly in any quantity which is normalized (i.e., O/H, \htwo/\mstar).
However, as before,
the different timescales for the different SF modes make the gas fraction 
relative to stars (and chemical abundance) non-unique tracers of evolution.
A low-mass  RE spheroid can achieve similar metallicity
and gas mass fraction as a more slowly evolving, but more massive, SE 
galaxy.
The gas fraction of the rapidly evolving spheroid will be higher than
that of the more slowly evolving  galaxy, 
as long as we observe the former at an earlier evolutionary phase
than the latter.

\begin{figure*}
\hbox{
\includegraphics[width=\linewidth,bb=18 144 592 718]{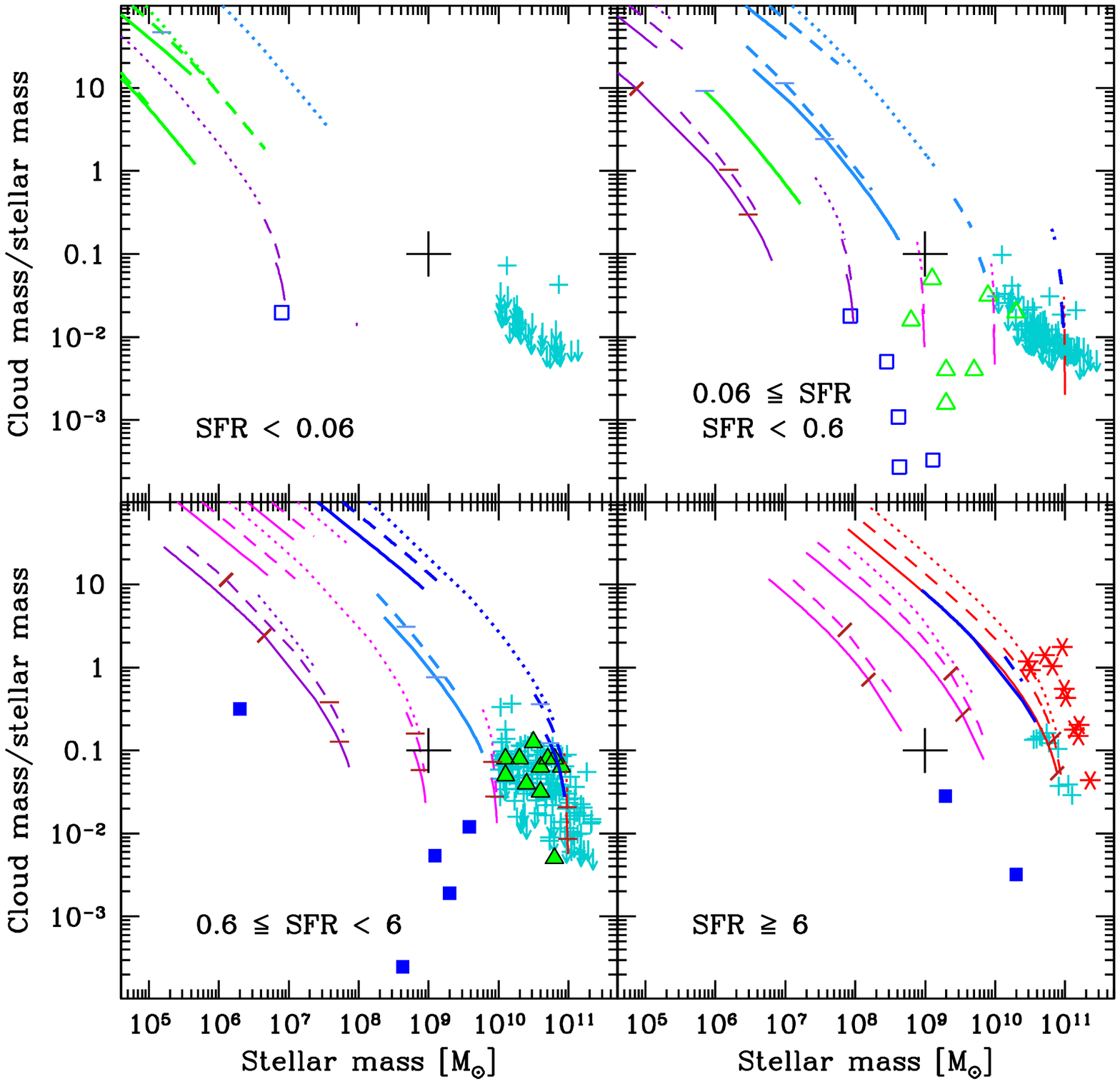} 
}
\caption{Ratio of molecular cloud mass and \mstar\ vs. stellar mass \mstar\
divided into regions of SFR.  The four regimes we consider are:
$\mbox{SFR}\leq 0.06$\,\msunyr, $0.06<\mbox{SFR}\leq 0.6$\,\msunyr, $0.6
<\mbox{SFR}\leq 6$\,\msunyr, $\mbox{SFR} \geq 6$\,\msunyr.  
Squares (blue) correspond to the BCDs from \citet{fumagalli10},
triangles (green) to the spiral galaxies from \citet{leroy08},
and 6-pronged asterisks to the SMGs from \citet{hainline11}.
Turquoise $+$ and down-arrows indicate the COLDGASS molecular
gas content from \citet{saintonge11a,saintonge11b}.
Models are coded by line type and color as in Fig.~\ref{fig:ssfr_mass_4p}.
The evolution times are marked along the tracks: 
an inclined tick indicates 100\,Myr (only for RE galaxies), 
and a horizontal tick 1\,Gyr. 
Time increases downward and to the right.
The large $+$ corresponds to a ``fiducial'' value of cloud mass/\mstar\ of
0.1 and \mstar\,=\,10$^9$\,\msun, intended to guide the eye to differences among
the panels.
}
\label{fig:gas_mass_4p}
\end{figure*}

In Fig.~\ref{fig:gas_mass_4p}
we have plotted the ratio of molecular cloud to stellar mass  vs. \mstar,
separated into four panels of SFR as in previous figures.
Also shown are several samples including the BCDs from \citet{fumagalli10}
(also considered in the previous analyses), 
spiral galaxies from \citet{leroy08}, galaxies from the 
COLDGASS survey by \citet{saintonge11a,saintonge11b}, 
and sub-millimetre galaxies (SMGs) 
at $z\sim2-3$ from \citet{hainline11}\footnote{There is only one sample in common
with the SFMS and MZR analysis, because of the current unavailability of molecular
gas observations in galaxies with measured stellar masses, SFR, and metallicities.}.
Conversion factors \xco\ to derive \htwo\ mass 
(``cloud mass'' in Fig.~\ref{fig:gas_mass_4p}) are given in the respective
papers, and vary with galaxy type.

The main problem with comparing molecular content of
metal-poor starbursts (lower left and lower right panels in 
Fig.~\ref{fig:gas_mass_4p}) lies with the
intrinsic difficulty of detecting molecules in the cool gas phase 
in a metal-poor ISM \citep[e.g.,][]{taylor98,barone00,leroy05}.
Moreover, the conversion of CO intensities to \htwo\ mass at low metallicities
is very uncertain
\citep[e.g.,][]{bolatto08,leroy09,wolfire10,leroy11}.
While \citet{bolatto08} find \xco\ not to vary
significantly with metal abundance in Local Group galaxies, at
metallicities \logoh$\simlt8.2$ there is
evidence for \xco\ factors of 10 to 20 times higher \citep{leroy09,leroy11}.
For the BCDs with CO observations in \citet{fumagalli10}, 
we have used a conservative value, $X_{\rm CO}\,=\,4\times10^{20}$\,cm$^{-2}$\,[K\,km\,s$^{-1}]^{-1}$,
\citep[e.g.,][]{bolatto08}; hence, the estimates
of \htwo\ mass in the metal-poor BCDs shown in Fig.~\ref{fig:gas_mass_4p}
could be highly underestimated (and in fact are outliers 
because of their low inferred cloud mass/stellar mass ratios).

The MICE model predictions of the ratio of molecular gas mass
to stellar mass shown in Fig.~\ref{fig:gas_mass_4p} are in fairly
good agreement with the observations, excepting the COLDGASS upper limits
in the upper left panel with SFR$<$0.06\,\msunyr.
This consistency is perhaps surprising, given that our models are
closed box, but strongly suggests that other parameters, such as dynamical
timescales resulting from compact size and high density, play an important,
if not primary, role in governing gas fractions.

Comparison with Figs.~\ref{fig:ssfr_mass_4p} and \ref{fig:oh_mass_4p} indicate
that the models that reproduce the scaling relations between SFR, stellar
mass, and O/H in the upper panels are relatively gas poor.
The upper left panel in Fig.~\ref{fig:ssfr_mass_4p} is not well reproduced by our models 
because the galaxies tend to be more massive than expected from the sSFR;
this is perhaps propagating into Fig.~\ref{fig:gas_mass_4p} where galaxies
are again too massive than would be predicted from their low SFR and low cloud mass fraction. 
The upper right panel is apparently occupied by either RE
RE galaxies or more slowly evolving galaxies which, in any case, are at the
end of their lifetimes, having already converted most of their gas into stars.
As the SFR increases (lower left panel), gas content
also increases, and for the most extreme case (lower right panel), the
mass ratio of molecular clouds to stars is $\sim$1, 
during the time when the galaxies also satisfy the scaling relations between
metallicity, stellar mass, and SFR.
The data for SMGs are entirely consistent with the models, and in fact have
observed gas-to-total mass fractions [clouds/(stars$+$clouds)] of 50\% or more \citep{tacconi06}.

With our MICE 
models we are able to link the gas content and the SFR
with the evolutionary timescales.  The galaxies we call {\em active} 
are observed close to the peak of the SFR with a still high gas
content, and are relatively chemically unevolved. During that phase of their evolution they
are still gas-rich {\em star-forming} galaxies but will rapidly move toward the
end of their SF phase.
In fact, their evolution can be reproduced using chemical evolution
models for RE (starburst) galaxies.  On the other hand, {\em passive}
galaxies are evolving at a very slow rate, so that their gas content 
and SFR will be relatively low during most of their lifetime. 

Recently, \citet{zahid12} presented a sample of metal-rich low-mass galaxies,
a galaxy population complementary to the metal-poor more massive starbursts 
discussed here.
They interpreted the relatively high metallicities in their sample as due to
the reduced gas content of the galaxies, such that the more metal-rich 
galaxies at a given stellar mass are more gas poor. 
In fact, the metal-rich low-mass galaxies studied by \citet{zahid12}
would occupy the upper panels in Fig.~\ref{fig:gas_mass_4p}.
The interpretation of \citet{zahid12} is that at a given stellar mass,
metal-rich galaxies have low gas fractions, and that the low gas fractions are
largely responsible for the high gas-phase oxygen abundance observed.  
Instead, our models suggest that the highest metallicities at a given mass
are not a consequence of the gas fraction, but that both gas fraction and 
metal abundance are consequences of evolutionary time scales, driven by the 
initial conditions in {\em passive} and {\em active} galaxies.
Thus, our models also appear to verify
the empirical conclusion reached by \citet{zahid12} that such 
galaxies are gas poor relative to their metal-poor counterparts at the 
same stellar mass.

\section{Discussion }\label{sec:discussion}

A significant fraction of galaxies in our sample
deviates from the MZR and SFMS in similar ways, independently of redshift. 
These galaxies are apparently too massive for their metallicity, or
alternatively, for their mass they are too metal-poor and
have a SFR that is too high. 
The common characteristic of these galaxies,
the low-metallicity starbursts, is the way they are selected from larger samples.
Locally, metal-poor starbursts tend to come from objective-prism, emission line surveys; 
they are objects with strong 
\hb, [\oiii$]\lambda$5007, or \ha\ lines, and thus moderate dust extinction
and high SFRs.
At $z\simgt1$, Lyman break selections, based on rest-frame UV colors, 
pick out star-forming galaxies with similar properties, namely
relatively low dust extinction and relatively high SFRs 
\citep[e.g.,][]{shapley11}.
Such galaxies also tend to be dominated 
by the most recent burst of star formation,
with times from burst of a few tens of Myr to $\simlt$1\,Gyr 
\citep{shapley05b,haberzettl12}.
At low redshift, this is a result of the selection of 
very high emission-line
equivalent widths either in \ha\ or in [\oiii] 
which implies that ionized gas makes a significant contribution to the total light
(implying a young age). 
At high redshift, this is a function of the color selection for strongly
star-forming galaxies that can be finely tuned to select
galaxy populations at different redshifts.
Hence, the low- and high-redshift LMSs, which deviate from
the SFMS and the MZR in similar ways, have similar properties
{\em because of the way they are selected}.
In fact, together they form a Fundamental Plane of O/H, SFR, and \mstar,
which is apparently independent of redshift.

The galaxies which deviate from the SFMS and MZR 
are well modeled by the {\em active} mode of star formation.
The starbursts in the lower right panels of Figs.~\ref{fig:ssfr_mass_4p} and
\ref{fig:oh_mass_4p} are observed during the extreme starburst
phase of their evolution which is rapid and intense, lasting up to $\simlt$200-500\,Myr.
On the other hand, the galaxies which follow the general trends in the scaling relations
form stars in the {\em passive} mode.
The models suggest that they are observed during a more advanced
long-lasting phase of their evolution,
which endures most of their lifetime.

What are the drivers of the {\em active} and {\em passive} modes of star formation?
There are two related aspects of our models which differentiate {\em active} and {\em
passive} SF modes:  their intrinsic characteristics in terms of size
(RE vs. SE) and/or density of their star forming regions
(diffuse, compact, and hyper-dense clouds), and their evolutionary age.
The evolutionary age indicates the time from the last main episode of
SF: for quiescent galaxies which have had an initial single main episode of SF, 
this might be comparable to a Hubble time, while for {\em active}
galaxies it might be the time elapsed after the last major episode of
SF.  Thus when we refer to {\em active} galaxies as young objects, it
means that they had their major episode of SF relatively recently,
up to several hundreds of Myr ago.

The predictions of our models are in agreement with recent analyses
of the structure of galaxies up to $z\sim1$ along the SFMS \citep{wuyts11}.
At all stellar masses and redshifts, they find that the largest disk-like galaxies
lie along the SFMS;
these would be the galaxies with the longest dynamical times.
The star-bursting outliers are more centrally concentrated (compact) than 
the galaxies along the main sequence; these would have shorter dynamical
times and thus be more rapidly evolving.
Thus, our models correctly predict
that compact RE {\em active} models reproduce the starburst
outliers of the SFMS, and that SE galaxies, with their larger sizes
and slower evolution, lie along it.

An {\em active} SF mode is more readily associated with mergers and interactions
than a {\em passive} one; in fact, the initial conditions we attribute to
active star formation may be induced by mergers and tidal compression.
However, mergers are not the only way to foster a
highly star-forming massive galaxy at $z\sim2$ \citep[e.g.,][]{genzel06}.
In fact, we hypothesize that the initial conditions for the {\em active} and {\em passive}
modes of SF are governed by stochastic processes, which can occur
even in isolation. 

Although our MICE models reproduce the scaling
relations of metallicity, stellar mass, and SFR rather nicely,
they have limitations; as already discussed, the main one is they do not take
into account either inflowing or outflowing gas. 
Other attempts to explain the MZR (and FMR) have been based on the
transfer of gas, i.e., accretion of metal-poor gas and outflow of
metal-enriched material \citep{mannucci10,yates12,dave12,dayal12}.
Recent theoretical work, such as the ``equilibrium approach'' of \citet{dave12},
consider stochastic rates of inflow and 
outflow that regulate SFR and the accumulation of stellar mass and metallicity.
Such models are quite successful in approximating the shape of the MZR
and its dependence on SFR, at least down to \logoh$\sim$8, SFR$\sim$1\,\msunyr, and 
\mstar$\sim10^{8.5}$\,\msun\ \citep{yates12}.
It is not clear, though, whether they are consistent
with the lower metallicities and higher SFRs in the deviant metal-poor starbursts. 

Another limitation of our models is the way we approximate the onset of
the molecular phase; the atomic gas must become
molecular and form clouds before it can fuel star formation.
Our models assume that molecular clouds form at a constant efficiency,
with a dependence on atomic gas mass, and are disrupted by
cloud-cloud collisions.
More complex models take into account the self-shielding effect of
dust on the UV interstellar radiation field \citep[e.g.,][]{krumholz09,krumholz11}.
Although our models are successful in predicting the trends of the SFMS 
and the MZR, and observed gas fractions as in Fig.~\ref{fig:gas_mass_4p},
they may be oversimplified.
A more sophisticated treatment of the transition to molecular gas
will be dealt with in a future paper (Schneider et al., in preparation).

Despite the limitations of the MICE models, they successfully predict
the observed trends in scaling relations over a wide range of O/H, SFR, and \mstar.
Their predictions are also in good agreement with observed molecular
gas mass fractions and SFR.
It is likely that some of the processes driving galaxy evolution are captured by
our models, namely the formation of a dominant molecular phase,
together with fast or slow evolutionary
timescales depending on stochastic initial conditions. 
It is also likely that the physics behind galaxy evolution
requires some gas exchange with the environment, and a regulation mechanism
that drives an equilibrium between inflow and outflow \citep[e.g.,][]{dave12}.
A combination of key features of both classes of models may be necessary
to understand the mechanisms which drive evolution of galaxies in the early
universe.

\section{Conclusions }\label{sec:conclusions}

From a sample of $\sim1100$ galaxies spanning a wide range of O/H, SFR, and \mstar,
we have defined a class of low-metallicity starbursts which deviate from the SFMS
and the MZR in the same way, independently of redshift.
The sample taken as a whole (although excluding the galaxies with 
\mstar$\geq3\times10^{10}$\,\msun) defines a Fundamental Plane whose orientation is defined
primarily by SFR and \mstar, and whose thickness is governed by O/H.
Our models show that this plane is populated by galaxies and by models
with different modes of star formation: an active starburst mode that is more common at high redshift,
and a more passive, quiescent mode that predominates in the local universe and
that defines the main trends in the scaling relations. 
Summarizing the results shown in 
Figs.~\ref{fig:ssfr_mass_4p}, 
\ref{fig:oh_mass_4p}, and \ref{fig:gas_mass_4p},
galaxies following the main trends in the SFMS and MZR form stars
in the {\em passive} mode. They are well modeled by
slowly evolving  galaxies with only moderately dense
molecular clouds (either ``diffuse'' or ``compact'' MCs) and a relatively low gas content.
These galaxies spend most of their life in this quiescent, passive,
evolutionary phase reaching asymptotically the conversion of their gas
into stars in more than a Hubble time. We find that observations
are consistent with the end-phases of the evolution of these galaxies predicted
by our models.  On the
other hand, the galaxies which deviate from the SFMS and MZR form
stars in the {\em active} mode, typical of gas-rich RE galaxies 
with masses from $10^{8}$ to $10^{11}$\,\msun,
or gas-rich SE galaxies with compact or hyper-dense clouds with
masses of $\sim10^{11}$\,\msun. 
In the extreme cases of very high SFR, the evolutionary phase is short, 
suggesting that we are observing these galaxies just
after the start of their most recent SF episode.

The existence of a Fundamental Plane and the success of our modeling of active/passive star
formation implies that the SFMS and MZR do not really change with lookback time, 
but rather that the galaxy populations that define them change with redshift.
The ``evolution'' in the scaling relations
is a result of selecting those galaxies which are most common
at a particular redshift. 
Equivalently, the scaling relations and the deviations from them
result from different modes of star formation: an active starburst mode that is
more common in the distant universe and a more
passive quiescent one that dominates nearby. 

\section*{Acknowledgments}
We are grateful to the 
International Space Science Institute (ISSI) for financial support for
our collaboration,
MODULO (MOlecules and DUst and LOw metallicity), so that we could meet and
discuss the work for this paper.
We warmly thank Ameli\'e Saintonge for providing us with the electronic version
of the COLDGASS data including SFRs.
We also thank the referee, Mercedes Moll\'a, for insightful comments and
suggestions which improved the paper.
L.M. acknowledges the support of the ASI-INAF grant I/009/10/0,
and L.H. and R.S. the support of 2009 PRIN-INAF funding.

\label{lastpage}

\end{document}